\documentclass[DIV=15, bibliography=totoc]{scrartcl}

\usepackage[a4paper,margin=2.25cm,bottom=2.5cm]{geometry}
\usepackage{float}
\usepackage[export]{adjustbox}

\usepackage{xcolor} 
\definecolor{lightblue}{rgb}{0,0.5,1.0}
\definecolor{linkblue}{rgb}{0,0.1,0.6}
\definecolor{citegreen}{rgb}{0,0.4,0.0}%{0.1,0.5,0.4}%{0.125,0.6,0.5}
\definecolor{linkred}{rgb}{0.8,0,0.005}%{0.6,0,0.1}
\definecolor{mailviolet}{rgb}{0.3,0,0.35}%{0.6,0,0.1}
\definecolor{tumblue}{rgb}{0,0.396,0.741}
\definecolor{darkgreen}{rgb}{0,0.4,0} 
\definecolor{darkbrown}{rgb}{0.5, 0.396, 0.09}

\usepackage[hyperindex,colorlinks=true,linkcolor=linkblue,citecolor=citegreen,urlcolor=mailviolet,filecolor=linkred]{hyperref}
\usepackage[hyphenbreaks]{breakurl}
\usepackage{caption, subcaption}
\usepackage{booktabs}   
\usepackage{amsmath}
\usepackage{amsfonts}
\usepackage{lmodern}
\usepackage{graphicx}
\usepackage{listings}
\usepackage{todonotes}
\usepackage{placeins}
\usepackage{fancyhdr}
\usepackage{wrapfig}

\usepackage{siunitx}
\usepackage[square,comma,nonamebreak,compress,numbers]{natbib}

\usepackage{soulpos}
\usepackage{ulem}
\usepackage{authblk} % used for affiliations 

\usepackage{pgfplots}
\usepackage{pgfplotstable}
\pgfplotsset{compat = newest}
\usepackage{filecontents}
\usepackage{tikz}
\usepackage{tikzscale}
\usepackage{tikz-3dplot}

\usetikzlibrary{spy}
\usetikzlibrary{intersections}
\usetikzlibrary{arrows,shapes}
\usetikzlibrary{spy}
\usetikzlibrary{backgrounds}  
\usetikzlibrary{decorations}
\usetikzlibrary{decorations.markings}
\usetikzlibrary{positioning}
\usetikzlibrary{patterns}
\usetikzlibrary{calc} 
\usetikzlibrary{quotes} 
\usetikzlibrary{external} % Is activated with the command '\tikzexternalize'.
\usetikzlibrary{matrix}

\pgfplotscreateplotcyclelist{customCycleList}
{%
  {blue, mark=o},
  {red, mark=square},
  {darkgreen, mark=diamond},
  {cyan, mark=diamond},
  {darkbrown, mark=triangle*},
  {black, mark=pentagon},
}

\pgfplotsset{every axis/.append style= {
    cycle list name=customCycleList,
}}

\title{Immersed boundary parametrizations for full waveform inversion}

\author[1]{Tim B\"urchner\thanks{\href{mailto:tim.buerchner@tum.de}{\texttt{tim.buerchner@tum.de}},
		Corresponding author}}
\author[1]{Philipp Kopp}
\author[1]{Stefan Kollmannsberger}
\author[1, 2]{Ernst Rank}

\affil[1]{Chair of Computational Modeling and Simulation, 
	Technische Universit\"at M\"unchen %\authorcr
}%,Arcisstr. 21, 80333 M\"unchen, Germany}

\affil[2]{Institute for Advanced Study, 
	Technische Universit\"at M\"unchen %\authorcr
}%,Lichtenbergstr. 2a, 85748 Garching, Germany}

\date{}
\fancypagestyle{plain}{%
	\fancyhf{}%
	\fancyfoot[L]{
		\vspace{-0.0cm} 
} }

\newcommand{\tensor}[1]{\mathbf{#1}}
\newcommand{\tabref}[1]{Table~\ref{#1}}
\newcommand{\figref}[1]{Figure~\ref{#1}}
\newcommand{\secref}[1]{Section~\ref{#1}}
\newcommand{\appref}[1]{Appendix~\ref{#1}}
\usepackage{xcolor}
\usepackage{comment} 

\begin{document}

\normalem \maketitle
\normalfont\fontsize{11}{13}\selectfont

% TODO:
% * 

% ---------- Abstract ----------
\vspace{-1.5cm} \hrule 

\section*{Abstract}

Full Waveform Inversion (FWI) is a successful and well-established inverse method for reconstructing material models from measured wave signals. In the field of seismic exploration, FWI has proven particularly successful in the reconstruction of smoothly varying material deviations. In contrast, non-destructive testing (NDT) often requires the detection and specification of sharp defects in a specimen. If the contrast between materials is low, FWI can be successfully applied to these problems as well. However, so far the method is not fully suitable to image defects such as voids, which are characterized by a high contrast in the material parameters. In this paper, we introduce a dimensionless scaling function $\gamma$ to model voids in the forward and inverse scalar wave equation problem. Depending on which material parameters this function $\gamma$ scales, different modeling approaches are presented, leading to three formulations of mono-parameter FWI and one formulation of two-parameter FWI. The resulting problems are solved by first-order optimization, where the gradient is computed by an ajdoint state method. The corresponding Fr\'echet kernels are derived for each approach and the associated minimization is performed using an L-BFGS algorithm. A comparison between the different approaches shows that scaling the density with $\gamma$ is most promising for parameterizing voids in the forward and inverse problem. Finally, in order to consider arbitrary complex geometries known a priori, this approach is combined with an immersed boundary method, the finite cell method (FCM). 

% ---------- Keywords ----------
\vspace{0.25cm}
\noindent \textit{Keywords:} 
full waveform inversion, scalar wave equation, adjoint state method, finite cell method, gradient-based optimization
\vspace{0.25cm}

% Content 

% ---------- Sections ----------
%\input{01_introduction}
\section{Introduction}
{
\label{sec:introduction}
FWI is a seismic imaging technique first introduced by Lailly in 1983 \cite{Lailly1983} and Tarantola in 1984~\cite{Tarantola1984}. Information about the internal structure of a medium is carried in waveforms that propagate from mechanical sources to receiver positions. Fitting simulated wavefields of an iteratively adapted numerical model to wavefields recorded in experiments reveals this information. Mathematically, the process of data-fitting is described by a nonlinear optimization problem. The material parameters are the optimization variables. Due to the high dimensionality and expensive objective function, gradient-based optimization algorithms are used to find an approximate minimum of the inverse problem. The gradient is computed using an adjoint state method, see \cite{Givoli2021}. In addition to the forward problem, an adjoint backward problem must be solved in each minimization step. Overviews of the state of the art for the application of FWI in geophysics can be found in \cite{Fichtner2011, Virieux2009, Vigh2008}. \figref{methodik} shows the general structure of an FWI algorithm.

Over the years, FWI has been successfully applied in many scientific fields. For medical application, Pratt et al. \cite{Pratt2007} used visco-acoustic FWI to reconstruct the wave speed and attenuation in in-vivo breast tissue. Later, Sandhu et al. extended this approach for 3D applications ~\cite{Sandhu2015, Sandhu2017}. In 2020, Guasch et al. obtained images of a human brain \cite{Guasch2020}. FWI also found favor in non-destructive testing (NDT) for industrial and civil engineering applications. In particular, ultrasonic testing of metal components offers a wide range of applications. Guided wave tomography can determine the thickness of walls due to corrosion and other damage by reconstructing the wave speed in plates or pipes \cite{Rao2016, Rao2017, Lin2020}. The approaches are based on 2D mono-parameter FWI for the acoustic wave equation. Further attempts have been made -- for example, to identify the position and size of reinforcements in concrete using 3D acoustic FWI \cite{Seidl2018} or to detect objects inside underwater structures \cite{Belykh2020}. Rao et al. proposed a multi-parameter approach for the simultaneous reconstruction of density and wave speed of material inclusions \cite{Rao2020}.
\begin{figure}[H]
	\centering
	\resizebox{0.75\textwidth}{!}{
		\begin{tikzpicture}
	\centering
	\filldraw[draw=black,fill=cyan, line width=0.5mm] (-0.075\textwidth,0.5\textwidth) rectangle (0.075\textwidth,0.45\textwidth);
	\node at (0.0\textwidth,0.475\textwidth) {\footnotesize Initial model};
	
	\draw[draw=black, ->, line width=0.5mm] (0.0\textwidth,0.45\textwidth) -- (0.0\textwidth,0.425\textwidth);
	
	\filldraw[draw=black,fill=white, line width=0.5mm] (-0.075\textwidth,0.375\textwidth) rectangle (0.075\textwidth,0.425\textwidth);
	\node at (0.0\textwidth,0.4\textwidth) {\footnotesize Current model};
		
	\draw[draw=black, ->, line width=0.5mm] (0.0\textwidth,0.375\textwidth) -- (0.0\textwidth,0.35\textwidth);
	
	\filldraw[draw=black,fill=white, line width=0.5mm] (-0.1\textwidth,0.35\textwidth) rectangle (0.1\textwidth,0.3\textwidth);
	\node at (0.0\textwidth,0.325\textwidth) {\footnotesize Forward simulation};
		
	\filldraw[draw=black,fill=white, line width=0.5mm] (0.42\textwidth,0.5\textwidth) rectangle (0.58\textwidth,0.45\textwidth);
	\node at (0.5\textwidth,0.475\textwidth) {\footnotesize Real structure};
	
	\draw[draw=black, ->, line width=0.5mm] (0.5\textwidth,0.45\textwidth) -- (0.5\textwidth,0.35\textwidth);
	
	\filldraw[draw=black,fill=white, line width=0.5mm] (0.4\textwidth,0.35\textwidth) rectangle (0.6\textwidth,0.3\textwidth);
	\node at (0.5\textwidth,0.325\textwidth) {\footnotesize Experimental data};
	
	\filldraw[draw=black, fill=white, line width=0.4mm] (0.25\textwidth,0.325\textwidth) circle (0.015\textwidth);
	\node at (0.25\textwidth,0.325\textwidth) {$-$};
	
	\draw[draw=black, ->, line width=0.5mm] (0.1\textwidth,0.325\textwidth) -- (0.235\textwidth,0.325\textwidth);
	\draw[draw=black, <-, line width=0.5mm] (0.265\textwidth,0.325\textwidth) -- (0.4\textwidth,0.325\textwidth);
	
	\draw[draw=black, ->, line width=0.5mm] (0.25\textwidth,0.31\textwidth) -- (0.25\textwidth,0.275\textwidth);
	
	\filldraw[draw=black,fill=white, line width=0.5mm] (0.175\textwidth,0.225\textwidth) rectangle (0.325\textwidth,0.275\textwidth);
	\node at (0.25\textwidth,0.25\textwidth) {\footnotesize Residual};
	
	\draw[draw=black, ->, line width=0.5mm] (0.25\textwidth,0.225\textwidth) -- (0.25\textwidth,0.2\textwidth);
	
	\filldraw[draw=black,fill=white, line width=0.5mm] (0.15\textwidth,0.2\textwidth) rectangle (0.35\textwidth,0.15\textwidth);
	\node at (0.25\textwidth,0.175\textwidth) {\footnotesize Backward simulation};
	
	\draw[draw=black, ->, line width=0.5mm] (0.25\textwidth,0.15\textwidth) -- (0.25\textwidth,0.125\textwidth);
	
	\filldraw[draw=black,fill=white, line width=0.5mm] (0.175\textwidth,0.075\textwidth) rectangle (0.325\textwidth,0.125\textwidth);
	\node at (0.25\textwidth,0.1\textwidth) {\footnotesize Gradient};
	
	\draw[draw=black, ->, line width=0.5mm] (0.25\textwidth,0.075\textwidth) -- (0.25\textwidth,0.05\textwidth);
	
	\filldraw[draw=black,fill=cyan, line width=0.5mm] (0.175\textwidth,0.05\textwidth) rectangle (0.325\textwidth,0.0\textwidth);
	\node at (0.25\textwidth,0.025\textwidth) {\footnotesize Model update};
	
	\draw[draw=black, ->, line width=0.5mm] (0.325\textwidth,0.025\textwidth) -- (0.425\textwidth,0.025\textwidth);
	
	\filldraw[draw=black,fill=green, line width=0.5mm] (0.425\textwidth,0.05\textwidth) rectangle (0.575\textwidth,0.0\textwidth);
	\node at (0.5\textwidth,0.025\textwidth) {\footnotesize Final model};
	
	\draw[draw=black, ->, line width=0.5mm] (-0.15\textwidth,0.4\textwidth) -- (-0.075\textwidth,0.4\textwidth);
	\draw[draw=black, line width=0.5mm] (-0.15\textwidth,0.025\textwidth) -- (-0.15\textwidth,0.4\textwidth);
	\draw[draw=black, line width=0.5mm] (-0.15\textwidth,0.025\textwidth) -- (0.175\textwidth,0.025\textwidth);
	
\end{tikzpicture}
	}
	\caption{Structure of an FWI algorithm after \cite{Seidl2018}}
	\label{methodik}
\end{figure}
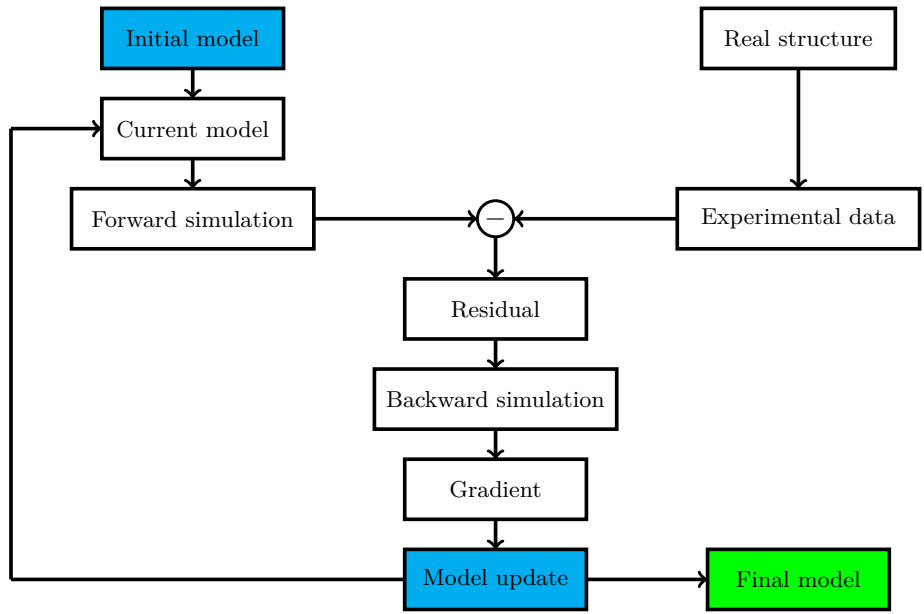%
From an experimental point of view, phased array ultrasonic tomography for internal defect detection is a rapidly growing technique in non-destructive testing. The phased array is attached to a single sample surface. One transducer at a time excites a wavelet that is measured by the other elements. The observed signals are stored in a full matrix capture (FMC). The total focusing method~(TFM), traveltime tomography (TT) and reverse time migration (RTM) are among the conventional methods for analyzing these data sets. Using these methods, several approaches have been developed to detect defects in concrete \cite{Ohara2021} and welds \cite{Cassels2018} or image cracks in metal components~\cite{Rao2019, Rao2021}. Reviews on this topic can be found in \cite{Drinkwater2006, Cleland2011, Felice2018}. While TFM, TT and RTM provide only a rough estimate of the defect's shape and a meaningful quantification of the material parameters is mostly not possible, a recently developed method using FWI is to directly invert the geometric shape of voids using boundary-conforming finite element meshes. Shi et al. reconstruct the $x$- and $y$-coordinates and the radius $r$ of an assumed circular void~\cite{Shi2019}. Sayag and Givoli generalize this approach to reconstruct more complex shapes by the inversion of a polygon~\cite{Sayag2022}. In both approaches, an initial guess of the void must be made.

Our contribution proposes an improvement of FWI in void modeling and reconstruction, as it does not require a priori assumptions about the number, shape and position of voids. In particular, we investigate the following three aspects: 
\vspace{-0.25cm}
\begin{itemize}
	\item First, we introduce different ways of parameterizing voids in the forward problem of the scalar wave equation and study the associated modeling errors. To this end, we introduce a function $\gamma$ that scales either only the density or the wave speed or both within the void region. The forward problem is solved for a 1D and 2D example.
	\vspace{-0.25cm}
	\item Second, the inverse problem is formulated with respect to the scaling function $\gamma$. The corresponding gradients are derived using the adjoint state method. Three different mono-parameter FWI formulations are used for the reconstruction of a void. For a comprehensive study, they are compared with the two-parameter FWI that simultaneously inverts the density and wave speed in one optimization. The comparison shows that scaling the density leads to the best void reconstruction.
	\vspace{-0.25cm}
	\item Finally, we combine the most promising approach to parameterize the inverse problem -- i.e., density scaling for void reconstruction -- with an immersed boundary method, the finite cell method~(FCM) \cite{Duester2017}. While a priori known geometric features are modeled by the FCM and its indicator function $\alpha$, unknown voids are detected and characterized by reconstructing the scaling function~$\gamma$.
\end{itemize}

The paper is organized as follows. \secref{sec:forwardproblem} presents possible parametrizations for modeling voids in the scalar wave equation forward problem. These formulations are analyzed in 1D and 2D settings. \secref{sec:inverseproblem} defines the inverse problem by describing the corresponding optimization problem. The gradient is computed using the adjoint state method. \secref{sec:circularvoid} compares the proposed approaches for the reconstruction of a circular void. In \secref{sec:ellipticalvoid}, we introduce the FCM to consider arbitrary complex geometries known a priori. The FWI approach is then combined with the FCM. As an example, an unknown elliptical void is reconstructed in an immersed boundary domain. \secref{sec:conclusion} gives a final conclusion.
}

\section{The forward problem}
{
\label{sec:forwardproblem}

\subsection{Scalar wave equation}

The time-domain 2D scalar wave equation in an isotropic heterogeneous media has the following form 
\begin{equation}
\rho(\tensor{x}) \ddot{u}(\tensor{x}, t) - \nabla \cdot \left( \rho(\tensor{x}) c^2(\tensor{x}) \nabla u(\tensor{x}, t) \right) = f(\tensor{x}, t)\text{,} \qquad \tensor{x} \in \Omega 
\label{PDE_scalar}
\end{equation}
in the domain of computation $\Omega$. In this equation, $u(\tensor{x}, t)$ and $\ddot{u}(\tensor{x}, t)$ are the solution field of the forward problem and its acceleration, $\rho(\tensor{x})$ and $c(\tensor{x})$ are the density and wave speed of the material and $f(\tensor{x}, t)$ is the external volume force. The initial conditions $u(\tensor{x}, 0) = 0$ and $\dot{u}(\tensor{x}, 0) = 0$, $\tensor{x} \in \Omega$ and boundary conditions $u(\tensor{x}, t) = 0\text{,} \, \tensor{x} \in \partial \Omega_\text{D}$ and $\tensor{n} \cdot \nabla u(\tensor{x},t) = 0\text{,} \, \tensor{x} \in \partial \Omega_\text{N}$ with $\partial \Omega = \partial \Omega_\text{D} \cup \partial \Omega_\text{N}$ are given. We assume that $\Omega$ is divided into a known intact part $\Omega_\text{I}$ and an unknown void part $\Omega_\text{V}$. To simplify our analysis, we also assume that the density and the wave speed take constant values $\rho_0$ and $c_0$ inside $\Omega_\text{I}$. We now need to find suitable values for $\rho$ and $c$ inside $\Omega_\text{V}$, such that we obtain an equivalent solution to solving \eqref{PDE_scalar} on $\Omega_\text{I}$. To this end, we introduce a scaling function $\gamma(\tensor{x})$ that is $1$ inside $\Omega_\text{I}$ and a small value close to $0$ inside $\Omega_\text{V}$. \figref{justFlaw} shows an example domain with an elliptical void inside an intact material block and the corresponding values for $\gamma$. We now introduce several ways of scaling $\rho$ and $c$ with $\gamma$ and investigate how suitable they are for modeling voids in a forward problem. In \secref{sec:inverseproblem}, we try to find $\Omega_V$ by formulating the inverse problem for reconstructing the function $\gamma$ that minimizes the error to some measurements.

\vspace{-0.4cm}
\begin{figure}[H]
	\centering
	\resizebox{0.8\textwidth}{!}{
		\input{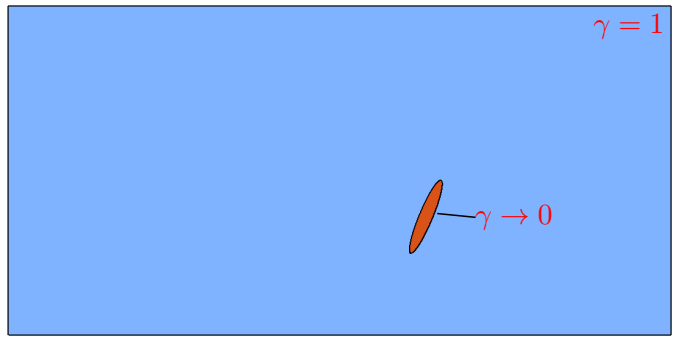}
	}
	\vspace{-1cm}
	\caption{Unknown void parameterized by the scaling function $\gamma$}
	\label{justFlaw}
\end{figure}

The product of density and wave speed is the acoustic impedance $I(\tensor{x}) = \rho(\tensor{x}) c(\tensor{x})$. At an interface between two materials, the difference in impedance determines the transmission and reflection of the incident wave. See \cite{Bedford1996} for detailed explanations. Physical voids (empty of material) are expected to have an impedance of $0$, which corresponds to reflection at a free boundary. Thus, one way to approximately parameterize voids is to apply the scaling function to the density $\rho(\tensor{x}) = \gamma(\tensor{x}) \rho_0$ while keeping the wave speed constant $c(\tensor{x}) = c_0$. A small scaling factor leads to impedances close to $0$ in the voids. Applying $\rho$-scaling leads to the scaled wave equation
\begin{equation}
\gamma(\tensor{x}) \rho_0 \ddot{u}(\tensor{x}, t) - \nabla \cdot \left( \gamma(\tensor{x}) \rho_0 c_0^2 \nabla u(\tensor{x}, t) \right) = f(\tensor{x}, t)\text{.}
\end{equation}

Scaling the wave speed $c(\tensor{x}) = \gamma(\tensor{x}) c_0$ with $\gamma(\tensor{x})$ at constant density $\rho(\tensor{x}) = \rho_0$ also leads to impedances near $0$ inside voids. With $c$-scaling, the scalar wave equation changes to
\begin{equation}
\rho_0 \ddot{u}(\tensor{x}, t) - \nabla \cdot \left( \rho_0 \gamma^2(\tensor{x}) c_0^2 \nabla u(\tensor{x}, t) \right) = f(\tensor{x}, t)\text{.}
\end{equation}

Third, it is possible to scale both parameters with the same scaling function $\gamma$. The wave speed is $c(\tensor{x}) = \gamma(\tensor{x}) c_0$ and the density $\rho(\tensor{x}) = \gamma(\tensor{x}) \rho_0$. For small $\gamma$, the $\rho c$-scaling also leads to impedances of almost $0$ within voids. The corresponding form of the scalar wave equation is
\begin{equation}
\gamma(\tensor{x}) \rho_0 (\tensor{x}) \ddot{u}(\tensor{x}, t) - \nabla \cdot \left( \gamma^3(\tensor{x}) \rho_0 c_0^2 (\tensor{x}) \nabla u(\tensor{x}, t) \right) = f(\tensor{x}, t)\text{.}
\end{equation}

Since the impedance is the product of density and wave speed, one may choose to scale both parameters separately by two independent scaling functions $\gamma_\rho$ and $\gamma_c$. They represent the relative density and wave speed with respect to the intact domain, $\rho(\tensor{x}) = \gamma_\rho(\tensor{x}) \rho_0$ and $c(\tensor{x}) = \gamma_c(\tensor{x}) c_0$. As long as one parameter $\gamma_\rho$ or $\gamma_c$ is almost $0$ inside the void regions, the impedance also decreases to nearly $0$. We call the parametrization separate-scaling and the scalar wave equation is
\begin{equation}
\gamma_\rho(\tensor{x}) \rho_0 \ddot{u}(\tensor{x}, t) - \nabla \cdot \left( \gamma_\rho(\tensor{x}) \rho_0 \gamma_c^2(\tensor{x}) c_0^2 \nabla u(\tensor{x}, t) \right) = f(\tensor{x}, t)\text{.}
\end{equation}

\subsection{Preliminary studies}
\label{preliminaryForward}

First, we investigate whether the scaling approaches presented above are capable of modeling interfaces between healthy material and voids using the finite element method. Since in inverse problems the position and size of the voids are not known in advance, we focus in particular on configurations where the element boundaries do \textit{not} coincide with the material interfaces. 

\subsubsection*{1D interface problem}

An example for a 1D interface problem is defined as follows. A domain $x \in \Omega = [0, l]$, of length $l = \SI{3}{\meter}$ is discretized with 1D finite elements of size $h_e = \SI{0.05}{\meter}$. Time integration is performed with second-order central differences. The first part of the domain $x \in [0, 2.0167]$ consists of intact homogeneous material with density $\rho_0 = \SI{1}{\kilogram \per \meter^3}$ and wave speed $c_0 = \SI{1}{\meter \per \second}$. At position $x = 2.0167$ inside the element $[2.0, 2.05]$ is an interface. The second part of the domain $x \in [2.0167, l]$ mimics void material and is modeled using the presented scaling approaches, i.e. $\rho$-, $c$- and $\rho c$-scaling. The scaling function is set to
\begin{equation}
	\gamma(x) = \begin{cases}
	1 \quad &\text{, }x \in [0, 2.0167] \\
	10^{-5} \quad &\text{, }x \in [2.0167, 3]
	\end{cases} \text{.}
\end{equation}
In the cut element the integrand of the element matrix is discontinuous. Therefore, an adapted integration technique needs to be applied. We use a composed integration by distributing the quadrature points on a binary tree of depth $5$ for the cut element $[2.0, 2.05]$, which has been successfully applied in the context of the finite cell method (see Section 5.1). A Gaussian bell curve with a central frequency at $f = \SI{1}{\hertz}$ is set as an initial condition at the left side of the domain. The propagation of the wave is simulated for $T = \SI{4}{\second}$ with a step size $\Delta t = \SI{1e-3}{\second}$. The three approaches are compared for finite element discretizations with polynomial degrees $p = 1, \,2,\, \text{and} \,\, 4$. In addition, the analytical solution for a wave reflected from a free boundary at $x = 2.0167$ is shown. The solutions for $p = 1$ are visualized in \figref{1D_p1}. The time stamps $t_1 = \SI{0.5}{\second}$, $t_2 = \SI{2.1}{\second}$ and $t = \SI{3.5}{\second}$ show the wave before hitting the interface, during hitting the interface and after reflection. The black dashed vertical lines at $x = 2.0$ and $x = 2.05$ visualize the boundaries of the cut element, and the red vertical line at $x = 2.0167$ indicates the material interface. For the polynomial degrees $p = 2$ and $p = 4$ the images are given in \appref{App:1D}.

It can be observed that the solutions from $\rho$-scaling and $\rho c$-scaling agree very well with the analytical solution at all three time stamps for $p = 1$. Slight shifts in time can be attributed to dispersion. Increasing the polynomial degree (Figures \ref{1D_p2} and \ref{1D_p4}) removes these inaccuracies and no differences to the analytical solution are visible. While in $\rho c$-scaling the void domain remains at rest after the wave hits the interface, in $\rho$-scaling a wave is transmitted. Since we are only interested in the solution in the physical domain, the solution in the void domain can be neglected as long as it does not affect the physical domain. With $c$-scaling, a large time delay is observed for $p=1$. Instead of improving the solution, increasing the polynomial degree even causes oscillations from the void region to return to the physical domain after the wave hits the interface, as Figures \ref{1D_p2} and \ref{1D_p4} show. 

\begin{figure}[H]
	\centering
	\begin{subfigure}[b]{1.0\textwidth}
		\centering
		\begin{tikzpicture}
	\begin{axis}[
		xmin = 0, xmax = 3,
		ymin = -0.5, ymax = 2.0,
		xtick distance = 0.5,
		ytick distance = 0.5,
		grid=both,
		width = \textwidth,
		height = 0.3\textwidth,
		xlabel = {$x$-coordinate},
		ylabel = {Displacement},
		legend style={at={(1,1)}, anchor=north east}]
		
		\addplot[red, line width=1.5pt] file[] {tikz/data_1Dforward/1Dforward_analytic_t1.dat};
		\addplot[black, dashed, line width=1.5pt] file[] {tikz/data_1Dforward/1Dforward_p1_v0_t1.dat};
		\addplot[blue, dash dot, line width=1.5pt] file[] {tikz/data_1Dforward/1Dforward_p1_v1_t1.dat};
		\addplot[green, dotted, line width=1.5pt] file[] {tikz/data_1Dforward/1Dforward_p1_v2_t1.dat};
		\addplot[black, dashed] file[] {tikz/data_1Dforward/elBoundary1.dat};
		\addplot[black, dashed] file[] {tikz/data_1Dforward/elBoundary2.dat};
		\addplot[red] file[] {tikz/data_1Dforward/interface.dat};
		
		\legend{analytic, $\rho$-scaling, $c$-scaling, $\rho c$-scaling}
	\end{axis}
\end{tikzpicture}
		\vspace{-0.5cm}
		\caption{$t = \SI{0.5}{\second}$}
	\end{subfigure}
	%\hfill
	\begin{subfigure}[b]{1.0\textwidth}
		\centering
		\begin{tikzpicture}
	\begin{axis}[
		xmin = 0, xmax = 3,
		ymin = -0.5, ymax = 2.0,
		xtick distance = 0.5,
		ytick distance = 0.5,
		grid=both,
		width = \textwidth,
		height = 0.3\textwidth,
		xlabel = {$x$-coordinate},
		ylabel = {Displacement},
		legend style={at={(1,1)}, anchor=north east}]
		
		\addplot[red, line width=1.5pt] file[] {tikz/data_1Dforward/1Dforward_analytic_t3.dat};
		\addplot[black, dashed, line width=1.5pt] file[] {tikz/data_1Dforward/1Dforward_p1_v0_t3.dat};
		\addplot[blue, dash dot, line width=1.5pt] file[] {tikz/data_1Dforward/1Dforward_p1_v1_t3.dat};
		\addplot[green, dotted, line width=1.5pt] file[] {tikz/data_1Dforward/1Dforward_p1_v2_t3.dat};
		\addplot[black, dashed] file[] {tikz/data_1Dforward/elBoundary1.dat};
		\addplot[black, dashed] file[] {tikz/data_1Dforward/elBoundary2.dat};
		\addplot[red] file[] {tikz/data_1Dforward/interface.dat};
		
		\legend{analytic, $\rho$-scaling, $c$-scaling, $\rho c$-scaling}
	\end{axis}
\end{tikzpicture}
		\vspace{-0.5cm}
		\caption{$t = \SI{2.1}{\second}$}
	\end{subfigure}
	%\hfill
	\begin{subfigure}[b]{1.0\textwidth}
		\centering
		\begin{tikzpicture}
	\begin{axis}[
		xmin = 0, xmax = 3,
		ymin = -0.5, ymax = 2.0,
		xtick distance = 0.5,
		ytick distance = 0.5,
		grid=both,
		width = \textwidth,
		height = 0.3\textwidth,
		xlabel = {$x$-coordinate},
		ylabel = {Displacement},
		legend style={at={(1,1)}, anchor=north east}]
		
		\addplot[red, line width=1.5pt] file[] {tikz/data_1Dforward/1Dforward_analytic_t4.dat};
		\addplot[black, dashed, line width=1.5pt] file[] {tikz/data_1Dforward/1Dforward_p1_v0_t4.dat};
		\addplot[blue, dash dot, line width=1.5pt] file[] {tikz/data_1Dforward/1Dforward_p1_v1_t4.dat};
		\addplot[green, dotted, line width=1.5pt] file[] {tikz/data_1Dforward/1Dforward_p1_v2_t4.dat};
		\addplot[black, dashed] file[] {tikz/data_1Dforward/elBoundary1.dat};
		\addplot[black, dashed] file[] {tikz/data_1Dforward/elBoundary2.dat};
		\addplot[red] file[] {tikz/data_1Dforward/interface.dat};
		
		\legend{analytic, $\rho$-scaling, $c$-scaling, $\rho c$-scaling}
	\end{axis}
\end{tikzpicture}
		\vspace{-0.5cm}
		\caption{$t = \SI{3.5}{\second}$}
	\end{subfigure}
	%hfill
	\caption{1D interface problem $p=1$}
	\label{1D_p1}
\end{figure}
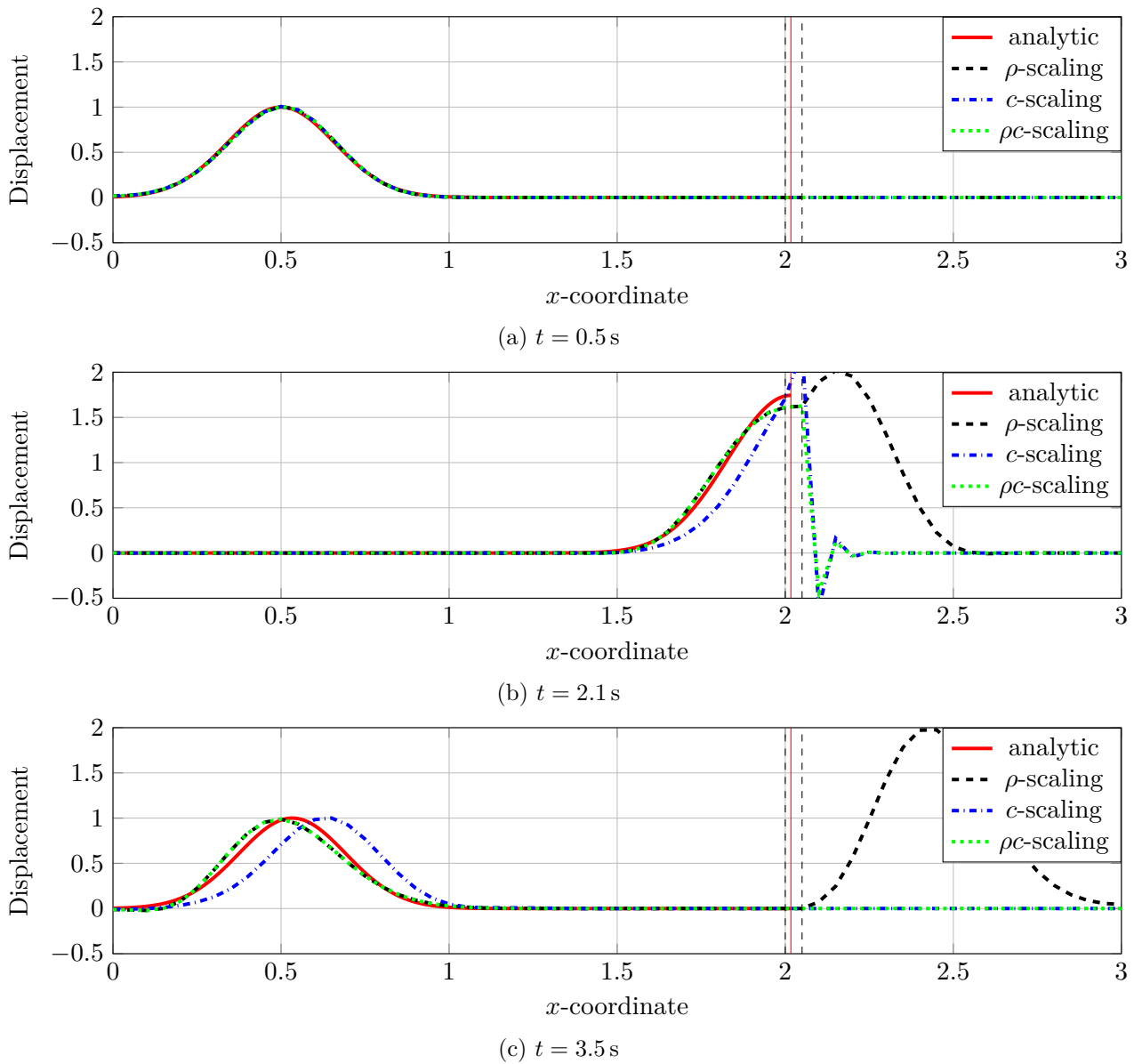

\subsubsection*{2D plate with circular void}

Our second forward example is a plate with a circular void defined by the following specifications. The plate of size $\SI{50}{\milli \meter} \times \SI{50}{\milli \meter}$ consists of a material with density $\rho_0 = \SI{2700}{\kilo / \meter^3}$ and wave speed $c_0 = \SI{6000}{\meter / \second}$. The domain is discretized with uniform square finite elements of size $h_e = \SI{0.5}{\milli \meter}$ leading to a mesh of $10 \, 000$ elements. Second-order central differences are used for time integration. The circular void with radius $r = \SI{5}{\milli \meter}$ in the center of the domain is modeled using $\rho$-, $c$- and $\rho c$-scaling. To incorporate the circular void, the scaling function is set to
\begin{equation}
	\gamma(\tensor{x}) = \begin{cases}
		10^{-5} \quad &\text{, }(x - 0.025)^2 + (y - 0.025)^2 \leq 0.005^2 \\
		1 \quad &\text{, elsewhere}		
	\end{cases} \text{.}
\end{equation}
Similar to the 1D case, a quadtree of depth $5$ is used to integrate the elements cut by the interface between healthy and void material. The plate is excited at the center of the top surface by a 2-cycle sine burst with a central frequency at $\SI{500}{\kilo \hertz}$. The propagation of the wave is simulated up to $T = \SI{1e-05}{\second}$ with a time step size $\Delta t = \SI{5e-10}{\second}$. Again, the approaches are compared for three finite elements discretizations with polynomial degrees $p = 1, \,2,\, \text{and} \,\,4$. The reference solution is computed using a boundary conforming unstructured mesh of $84 \, 144$ linear quadrilateral elements generated using Gmsh \cite{gmsh}. For $p = 1$, \figref{2D_p1} shows the solutions for all approaches at time $t = \SI{0.8e-05}{\second}$ after the wave hits the interface of the circular void. The dotted white circles visualize the interface. The solutions for polynomial degrees $p = 2$ and $p = 4$ are given in \appref{App:2D}. 

\begin{figure}[H]
	\centering
	%\hspace{0.05\textwidth}
	\begin{subfigure}[b]{0.4\textwidth}
		\centering
		\includegraphics[width=0.75\textwidth, left]{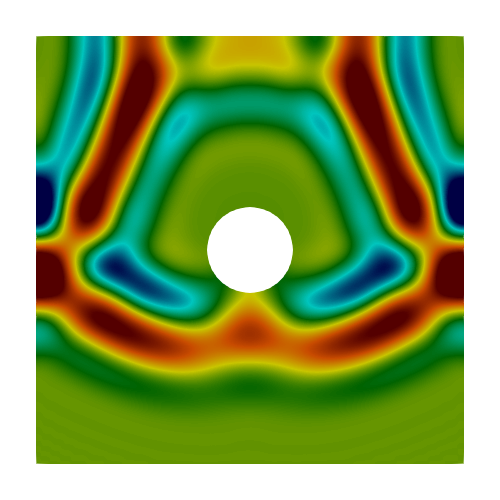}
		\vspace{-0.75cm}
		\captionsetup{margin={-0.25\textwidth,0cm}}
		\caption{reference}
	\end{subfigure}
	%\hspace{0.05\textwidth}
	%\hfill
	\begin{subfigure}[b]{0.3\textwidth}
		\centering
		\includegraphics[width=\textwidth]{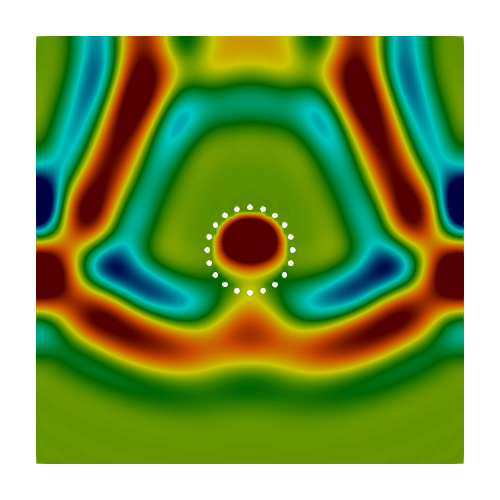}
		\vspace{-0.75cm}
		\caption{$\rho$-scaling}
	\end{subfigure}
	\begin{subfigure}[t]{0.07\textwidth}
		\centering		
		\includegraphics{tikz/Colorbar_-0p01to0p01.tikz}
	\end{subfigure}
	\hspace{3cm}
	%\hfill
	\begin{subfigure}[b]{0.4\textwidth}
		\centering
		%\resizebox{0.8\textwidth}{!}{
		\input{tikz/c_p1_withZoom.tex}
		%}
		%\includegraphics[width=\textwidth]{pics/c_p1.png}
		%\vspace{-0.75cm}
		\captionsetup{margin={-0.25\textwidth,0cm}}
		\caption{$c$-scaling}
	\end{subfigure}
	%\hfill
	\begin{subfigure}[b]{0.3\textwidth}
		\centering
		\includegraphics[width=\textwidth]{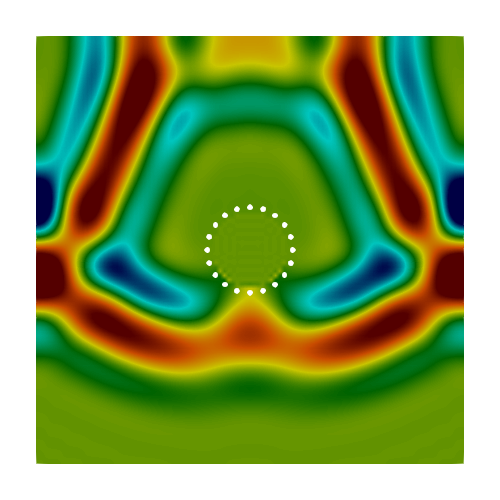}
		%\vspace{-0.75cm}
		\caption{$\rho c$-scaling}
	\end{subfigure}
	\begin{subfigure}[t]{0.07\textwidth}
		\centering		
		\includegraphics{tikz/Colorbar_-0p01to0p01.tikz}
	\end{subfigure}
	%hfill
	\caption{2D plate with void $p=1$ at time $t = \SI{0.8e-05}{\second}$}
	\label{2D_p1}
\end{figure}

Similar to the solutions for the 1D interface problem, the $\rho$-scaling and $\rho c$-scaling formulations lead to very good results. Neither for $p = 1$ nor for $p = 2$ or $p = 4$ is there much difference from the reference solution. Again, for $\rho$-scaling the wave is transmitted into the void, while for $\rho c$-scaling the void region is not excited. For both formulations, the solution in the physical domain is not visibly affected by the solution in the void region. By contrast, the application of $c$-scaling leads to oscillations in the physical domain after the wave hits the interface with the void. This disturbance can be explained by a 'boundary layer' of high oscillations inside the void, which influences the physical part of the domain via the cut elements. This boundary layer can be seen as 'colored dots' in the zoomed part of Figure 4c. In this 2D example, the solution for polynomial degree $1$ is disturbed the most. Also for higher polynomial degrees $p = 2$ and $p = 4$, these undesired perturbations do not vanish.

}

\section{The inverse problem}
{
\label{sec:inverseproblem}

\subsection{Optimization}

The goal of FWI is to find a material model $\tensor{m}^\ast(\tensor{x})$ that reproduces an experimental data set by numerical simulations, called forward simulations. The data set is acquired separately for $N_s$ experiments at the $N_r$ receiver positions. The material model is the set of spatially distributed material parameters $\tensor{m}(\tensor{x}) = \left[ m_1(\tensor{x}), m_2(\tensor{x}), m_3(\tensor{x}), ... \right]$ of the PDE. Mathematically, the FWI is described by a nonlinear optimization problem. To simplify the notation, the following equations are derived for a single source. For $N_s > 1$ the equations can be generalized by summing up the objective functions and gradients of the individual experiments. The objective function $\chi$ describes the misfit between measured and simulated wave signals. For a single source the optimization problem is defined as follows
\begin{equation}
\tensor{m}^\ast = \arg \min_{\tensor{m}} \chi(\tensor{m})
\end{equation}
where $\chi$ is the summed mean squared error of all receiver signals integrated over time $T$
\begin{equation}
\chi(\tensor{m}) = \frac{1}{2} \int_T \int_\Omega \chi_1(\tensor{m}) d \Omega dt = \frac{1}{2} \int_T \int_\Omega \sum_{r=1}^{N_r} \left[ \left( u(\tensor{m}; \tensor{x}, t) - u^0(\tensor{x}^r, t) \right)^2 \delta(\tensor{x} - \tensor{x^r}) \right] d \Omega dt
\end{equation}
and $\chi_1$ denotes the integrand of the space-time integral. At the receiver position $\tensor{x}^r$ the signal recorded over the measurement time is $u^0(\tensor{x}^r, t)$ and $u(\tensor{m}; \tensor{x}, t)$ denotes the simulated wavefield of the current model $\tensor{m}$. Due to the expensive forward problem and high dimensionality of the model space, gradient-based methods play a dominating role in solving the optimization problem. The discretized model vector~$\hat{\tensor{m}}$ is iteratively updated using a descent update step $\Delta \hat{\tensor{m}}$
\begin{equation}
\hat{\tensor{m}}^{(k+1)} = \hat{\tensor{m}}^{(k)} + \Delta \hat{\tensor{m}}^{(k)}\text{.}
\end{equation}
The superscript denotes the current iteration $k$. Quasi-Newton type optimization methods calculate the first derivative and estimate the second derivative of the objective function. The calculated gradient $\nabla_{\hat{\tensor{m}}} \chi (\hat{\tensor{m}}^{(k)})$ and approximated inverse Hessian $\tensor{H}_a^{-1} (\hat{\tensor{m}}^{(k)})$ define the model update
\begin{equation}
\Delta \hat{\tensor{m}}^{(k)} = - \tensor{H}_a^{-1} (\hat{\tensor{m}}^{(k)}) \nabla_{\hat{\tensor{m}}} \chi (\hat{\tensor{m}}^{(k)})\text{.}
\end{equation}

In this contribution, we use the limited-memory BFGS algorithm from the Python library SciPy~\cite{2020SciPy}. Instead of explicitly computing and storing the dense approximation of the inverse Hessian~$\tensor{H}_a^{-1} (\hat{\tensor{m}}^{(k)})$, the L-BFGS algorithm only stores the gradient and some auxiliary vectors from the last $M$ iterations. The algorithm is used to compute the descent direction in a matrix-free manner. For more information on gradient-based optimization, see \cite{Nocedal2006}.

\subsection{Gradient computation}

The adjoint state method originates from the optimal control theory. By introducing an adjoint state, the gradients in optimization problems can be computed efficiently. The derivation of the gradients in this paper closely follows Fichtner \cite{Fichtner2006a, Fichtner2006b}. Again, the procedure is presented for a single source to simplify notation during the derivation. For $N_s > 1$ sources, the gradients of the individual sources can be summed to form the final gradient. In the time domain, the derivative of the objective function with respect to a perturbation of the material model in direction $\delta \tensor{m}$ can be calculated as follows
\begin{equation}
\nabla_m \chi \delta \tensor{m} = \int_T \int_\Omega \tensor{u}^\dagger(\tensor{x}, t) \cdot \nabla_m \tensor{L}(\tensor{u}; \tensor{m}, \tensor{x}, t) \delta \tensor{m} d \Omega dt \label{generalAG}\text{.}
\end{equation}
where $\tensor{L}$ is the PDE operator, which describes the physics connecting the material parameters $\tensor{m}$ and the solution field $\tensor{u}$. The field $\tensor{u}^\dagger$ solves the adjoint equation. More information about the operator $L$ is given in~\cite{Fichtner2011}. For the scalar wave equation, the operator $L$ is linear in the solution field $u$ and describes the corresponding formulation of the PDE. For the mono-parameter FWI, our scaling function is the only inversion parameter $m(\tensor{x}) = \gamma(\tensor{x})$. The two-parameter approach has two inversion parameters $\tensor{m}(\tensor{x}) = \left[ \gamma_\rho(\tensor{x}), \gamma_c(\tensor{x}) \right]$. To simplify the notation, we omit the dependencies on $\tensor{x}$ and $t$ in the derivation the Fr\'echet kernels for the above parametrizations. The computation of the gradient with the adjoint state method is exemplarily derived for the $\rho$-scaling. The derivations for the other parametrizations can be performed analogously. The scalar operator $L$ is given by
\begin{equation}
L(u; \gamma) = \gamma \rho_0 \ddot{u} - \nabla \cdot \left( \gamma \rho_0 c_0^2 \nabla u \right)\text{.}
\end{equation}
Moreover, homogeneous initial conditions $u(\tensor{x}, t_0) = 0$ and $\dot{u}(\tensor{x}, t_0) = 0$ are given. Since the scalar wave-equation is self-adjoint for non-dissipative material, its adjoint equation is of the same type
\begin{equation}
\gamma \rho_0 \ddot{u}^\dagger - \nabla \cdot \left( \gamma \rho_0 c_0^2 \nabla u^\dagger \right) = f^\dagger\text{,} \qquad \tensor{x} \in \Omega \text{,}
\end{equation}
where the adjoint force results from the objective integrand $\chi_1$
\begin{equation}
f^\dagger = - \nabla_u \chi_1 = - \sum_{r=1}^{N_r} [u - u^0] \delta(\tensor{x} - \tensor{x}^r)\text{.}
\end{equation}
For the adjoint problem, homogeneous end conditions $u(\tensor{x}, t_{\text{end}}) = 0$ and $\dot{u}(\tensor{x}, t_{\text{end}}) = 0$ must be set. This can be interpreted as a time-inverted wave propagation, i.e. back-injecting the wave differences of an experiment at the receiver positions and propagating them backward in time. The boundary conditions $u(\tensor{x}, t) = 0\text{,} \, \tensor{x} \in \partial \Omega_\text{D}$ and $\tensor{n} \cdot \nabla u(\tensor{x},t) = 0\text{,} \, \tensor{x} \in \partial \Omega_\text{N}$ with $\partial \Omega = \partial \Omega_\text{D} \cup \partial \Omega_\text{N}$ remain the same as for the forward problem. The term $\nabla_\gamma L(u; \gamma)$ is calculated using the G\^ateaux derivative
\begin{equation}
\begin{split}
\nabla_\gamma L(u; v) &= \lim_{\epsilon \rightarrow \infty} \frac{1}{\epsilon} \left[ L(u; \gamma + \delta \gamma) - L(u; \gamma) \right] \\ &= \rho_0 \delta \gamma \ddot{u} - \cdot \nabla \left( \rho_0 c_0^2 \delta \gamma \nabla u \right)\text{.}
\end{split}
\end{equation}
After substituting $\nabla_\gamma L(u; \gamma)$ into \eqref{generalAG} and integrating by parts, the derivative is given by
\begin{equation}
\nabla_\gamma \chi \delta \gamma = \int_\Omega \underbrace{\int_T \left[ - \rho_0 \dot{u}^\dagger \dot{u} + \rho_0 c_0^2 \nabla u^\dagger \cdot \nabla u \right] dt}_{= K_\gamma} \delta \gamma d \Omega\text{,}
\end{equation}
where $K_\gamma$ is called the Fr\'echet kernel. The boundary terms of the integration by parts vanish due to the underlying initial and boundary conditions of the forward and adjoint problems. The different Fr\'echet kernels for all three formulations of mono-parameter FWI and for the two-parameter approach are shown in \tabref{FrechetTable}.
\begin{table}[H]
	\centering
	\begin{tabular}{ | l | l | }
		\hline
		\rule{0pt}{4ex} Formulation & Fr\'echet kernel \\
		\hline \hline 
		\rule{0pt}{4ex} $\rho$-scaling & $K_\gamma = \int_T \left[ - \rho_0 \dot{u}^\dagger \dot{u} + \rho_0 c_0^2 \nabla u^\dagger \cdot \nabla u \right] dt$ \\
		\hline
		\rule{0pt}{4ex} $c$-scaling & $K_\gamma = \int_T \left[ 2 \rho_0 c_0^2 \gamma \nabla u^\dagger \cdot \nabla u \right] dt$ \\
		\hline
		\rule{0pt}{4ex} $\rho c$-scaling & $K_\gamma = \int_T \left[ - \rho_0 \dot{u}^\dagger \dot{u} + 3 \rho_0 c_0^2 \gamma^2 \nabla u^\dagger \cdot \nabla u \right] dt$ \\
		\hline
		\rule{0pt}{4ex} separate-scaling & $K_{\gamma_\rho} = \int_T \left[ - \rho_0 \dot{u}^\dagger \dot{u} + \rho_0 c_0^2 \gamma_c^2 \nabla u^\dagger \cdot \nabla u \right] dt$ \\
		\rule{0pt}{4ex} & $K_{\gamma_c} = \int_T \left[ 2 \rho_0 c_0^2 \gamma_\rho \gamma_c \nabla u^\dagger \cdot \nabla u \right] dt$ \\
		\hline
	\end{tabular}
	\caption{Fr\'echet kernels}
	\label{FrechetTable}
\end{table}

Using the finite element method for spatial discretization, the solution field and adjoint field are discretized with $N_b$ basis functions $N_i$
\begin{subequations}
	\begin{align}
	u(\tensor{x},t) \approx \tilde{u}(\tensor{x},t) &= \sum_{i=1}^{N_b} N_i(\tensor{x}) \hat{u}_i(t) = \tensor{N}(\tensor{x}) \hat{\tensor{u}}(t)\text{,} \\ 
	u^\dagger(\tensor{x},t) \approx \tilde{u}^\dagger(\tensor{x},t) &= \sum_{i=1}^{N_b} N_i(\tensor{x}) \hat{u}_i^\dagger(t) = \tensor{N}(\tensor{x}) \hat{\tensor{u}}^\dagger(t)\text{.}
	\end{align}
\end{subequations}
In our approach, the same basis functions are used for the discretization of the material. Thus, the scaling function is represented by
\begin{equation}
\gamma(\tensor{x}) \approx \tilde{\gamma}(\tensor{x}) = \sum_{i=1}^{N_b} N_i(\tensor{x}) \hat{\gamma}_i = \tensor{N}(\tensor{x}) \hat{\tensor{\gamma}}\text{.}
\end{equation}
The derivative of the objective function with respect to the material coefficients $\hat{\gamma}_i$ can be calculated by computing the sensitivity with respect to the corresponding basis functions:
\begin{equation}
\frac{\partial \chi}{\partial \hat{\gamma}_i} = \nabla_\gamma \chi N_i = \int_\Omega K_\gamma N_i d \Omega\text{,}
\end{equation}
or in vector form
\begin{equation}
\nabla_{\hat{\tensor{\gamma}}} \chi = \nabla_\gamma \chi \tensor{N} = \int_\Omega K_\gamma \tensor{N} d \Omega\text{.}
\end{equation}
The discretized gradient vectors for all formulations are shown in \tabref{DiscretizedGradient}.
\begin{table}[H]
	\centering
	\begin{tabular}{ | l | l | }
		\hline
		\rule{0pt}{4ex} Formulation & Gradient vector \\
		\hline \hline 
		\rule{0pt}{4ex} $\rho$-scaling & $\nabla_{\hat{\tensor{\gamma}}} \chi = \int_\Omega \int_T \left[ - \rho_0 (\dot{\hat{\tensor{u}}}^\dagger)^T \tensor{N}^T \tensor{N} \dot{\hat{\tensor{u}}} + \rho_0 c_0^2 (\hat{\tensor{u}}^\dagger)^T \tensor{B}^T \tensor{B} \hat{\tensor{u}} \right] dt \tensor{N} d \Omega$ \\
		\hline
		\rule{0pt}{4ex} $c$-scaling & $\nabla_{\hat{\tensor{\gamma}}} \chi = \int_\Omega \int_T \left[ 2 \rho_0 c_0^2 \tensor{N} \hat{\tensor{\gamma}} (\hat{\tensor{u}}^\dagger)^T \tensor{B}^T \tensor{B} \hat{\tensor{u}} \right] dt \tensor{N} d \Omega$ \\
		\hline
		\rule{0pt}{4ex} $\rho c$-scaling & $\nabla_{\hat{\tensor{\gamma}}} \chi = \int_\Omega \int_T \left[ - \rho_0 (\dot{\hat{\tensor{u}}}^\dagger)^T \tensor{N}^T \tensor{N} \dot{\hat{\tensor{u}}} + 3 \rho_0 c_0^2 (\tensor{N} \hat{\tensor{\gamma}})^2 (\hat{\tensor{u}}^\dagger)^T \tensor{B}^T \tensor{B} \hat{\tensor{u}} \right] dt \tensor{N} d \Omega$ \\
		\hline
		\rule{0pt}{4ex} separate-scaling & $\nabla_{\hat{\tensor{\gamma}}_\rho} \chi = \int_\Omega \int_T \left[ - \rho_0 (\dot{\hat{\tensor{u}}}^\dagger)^T \tensor{N}^T \tensor{N} \dot{\hat{\tensor{u}}} + \rho_0 c_0^2 (\tensor{N} \hat{\tensor{\gamma}}_c)^2 (\hat{\tensor{u}}^\dagger)^T \tensor{B}^T \tensor{B} \hat{\tensor{u}} \right] dt \tensor{N} d \Omega$ \\
		\rule{0pt}{4ex} & $\nabla_{\hat{\tensor{\gamma}}_c} \chi = \int_\Omega \int_T \left[ 2 \rho_0 c_0^2 \tensor{N} \hat{\tensor{\gamma}}_\rho \tensor{N} \hat{\tensor{\gamma}}_c (\hat{\tensor{u}}^\dagger)^T \tensor{B}^T \tensor{B} \hat{\tensor{u}} \right] dt \tensor{N} d \Omega$ \\
		\hline
	\end{tabular}
	\caption{Discretized gradient vectors}
	\label{DiscretizedGradient}
\end{table}

}

\section{Plate with circular void}
{
\label{sec:circularvoid}

\subsection{Configuration}

In this section, we compare the presented FWI formulations in terms of their ability to reconstruct a circular void in a homogeneous intact material. For this purpose, a synthetic reference data set is generated by simulations of the structure with the void using a boundary conforming mesh. The size of the domain is $\SI{100}{\milli \meter} \times \SI{50}{\milli \meter}$. The circular void is located in the center of the domain and has a diameter $d = \SI{5}{\milli \meter}$. \figref{specimen_circle} shows the 'flawed' sample. The density of the material is set to $\rho = \SI{2700}{\kilogram \per \meter^3}$ and the wave speed to $\SI{6000}{\meter \per \second}$. The measurement data are computed with the 2D scalar wave equation on a boundary conforming unstructured mesh of $105\,459$ linear quadrilateral elements generated using Gmsh. A phased array, centrally located on the top surface of the sample, excites and measures the wave signals. The array consists of 65 equally spaced transducers with a pitch of $\SI{1}{\milli \meter}$. The excitation signal is a 2-cycle sine burst with a central frequency at $\SI{500}{\kilo \hertz}$. The propagation of the waves is simulated by second-order central differences for a period of $\SI{6.0 e-5}{\second}$ with a time step size of $\Delta t = \SI{5.0e-09}{\second}$ for the reference model and $\Delta t = \SI{1.0e-08}{\second}$ for the inversion model. The inversion model has a coarser uniform mesh with $20\,000$ linear quadrilateral elements of size $\SI{0.5}{\milli \meter} \times \SI{0.5}{\milli \meter}$. The same basis functions used for the solution field discretize the scaling functions $\gamma$ for mono-parameter FWI and $\gamma_\rho$ and $\gamma_c$ for two-parameter FWI. No further regularization terms are added. The scaling functions are clipped between $0.0$ and $1.2$ during the inversions. For each approach $25$~L-BFGS-B iterations are performed.

\begin{figure}[H]
	\centering
	\resizebox{0.6\textwidth}{!}{
		\begin{tikzpicture}
	\centering
	\filldraw[draw=gray,fill=white, line width=0.4mm] (0.108\textwidth,0.3\textwidth) rectangle (0.492\textwidth,0.33\textwidth);
	\draw [draw=gray, line width=0.1mm] (0.108\textwidth,0.33\textwidth) -- (0.108\textwidth,0.38\textwidth);
	\draw [draw=gray, line width=0.1mm] (0.492\textwidth,0.33\textwidth) -- (0.492\textwidth,0.38\textwidth);
	\draw[draw=gray, <->] (0.108\textwidth,0.355\textwidth) -- (0.492\textwidth,0.355\textwidth);
	\node at (0.3\textwidth,0.38\textwidth) {$\SI{64}{\milli \meter}$};
	\draw [draw=gray, line width=0.1mm] (0.132\textwidth,0.3\textwidth) -- (0.132\textwidth, 0.33\textwidth);
	\draw [draw=gray, line width=0.1mm] (0.156\textwidth,0.3\textwidth) -- (0.156\textwidth, 0.33\textwidth);
	\draw [draw=gray, line width=0.1mm] (0.180\textwidth,0.3\textwidth) -- (0.180\textwidth, 0.33\textwidth);
	\draw [draw=gray, line width=0.1mm] (0.204\textwidth,0.3\textwidth) -- (0.204\textwidth, 0.33\textwidth);
	\draw [draw=gray, line width=0.1mm] (0.228\textwidth,0.3\textwidth) -- (0.228\textwidth, 0.33\textwidth);
	\draw [draw=gray, line width=0.1mm] (0.252\textwidth,0.3\textwidth) -- (0.252\textwidth, 0.33\textwidth);
	\draw [draw=gray, line width=0.1mm] (0.276\textwidth,0.3\textwidth) -- (0.276\textwidth, 0.33\textwidth);
	\draw [draw=gray, line width=0.1mm] (0.300\textwidth,0.3\textwidth) -- (0.300\textwidth, 0.33\textwidth);
	\draw [draw=gray, line width=0.1mm] (0.324\textwidth,0.3\textwidth) -- (0.324\textwidth, 0.33\textwidth);
	\draw [draw=gray, line width=0.1mm] (0.348\textwidth,0.3\textwidth) -- (0.348\textwidth, 0.33\textwidth);
	\draw [draw=gray, line width=0.1mm] (0.372\textwidth,0.3\textwidth) -- (0.372\textwidth, 0.33\textwidth);
	\draw [draw=gray, line width=0.1mm] (0.396\textwidth,0.3\textwidth) -- (0.396\textwidth, 0.33\textwidth);
	\draw [draw=gray, line width=0.1mm] (0.420\textwidth,0.3\textwidth) -- (0.420\textwidth, 0.33\textwidth);
	\draw [draw=gray, line width=0.1mm] (0.444\textwidth,0.3\textwidth) -- (0.444\textwidth, 0.33\textwidth);
	\draw [draw=gray, line width=0.1mm] (0.468\textwidth,0.3\textwidth) -- (0.468\textwidth, 0.33\textwidth);
	\filldraw[draw=black,fill=white, line width=0.8mm] (0.0,0.0) rectangle (0.6\textwidth,0.3\textwidth);
	\draw [draw=gray, line width=0.1mm] (0.6\textwidth,0.0) -- (0.65\textwidth, 0.0);
	\draw [draw=gray, line width=0.1mm] (0.6\textwidth,0.3\textwidth) -- (0.65\textwidth,
	0.3\textwidth);
	\draw[draw=gray, <->] (0.625\textwidth,0.0) -- (0.625\textwidth,0.3\textwidth);
	\node at (0.675\textwidth,0.15\textwidth) {$\SI{50}{\milli \meter}$};
	\draw [draw=gray, line width=0.1mm] (0.0,0.0) -- (0.0, -0.05\textwidth);
	\draw [draw=gray, line width=0.1mm] (0.6\textwidth,0.0) -- (0.6\textwidth, -0.05\textwidth);
	\draw[draw=gray, <->] (0.0,-0.025\textwidth) -- (0.6\textwidth,-0.025\textwidth);
	\node at (0.3\textwidth,-0.05\textwidth) {$\SI{100}{\milli \meter}$};
	\filldraw[draw=black,fill=white, line width=0.4mm] (0.3\textwidth,0.15\textwidth) circle (0.015\textwidth);
	\draw [draw=gray, line width=0.1mm] (0.3\textwidth,0.0) -- (0.35\textwidth, 0.0);
	\draw [draw=gray, line width=0.1mm] (0.3\textwidth,0.135\textwidth) -- (0.35\textwidth, 0.135\textwidth);
	\draw [draw=gray, line width=0.1mm] (0.3\textwidth,0.165\textwidth) -- (0.35\textwidth, 0.165\textwidth);
	\draw[draw=gray, <->] (0.335\textwidth,0.0) -- (0.335\textwidth,0.135\textwidth);
	\draw[draw=gray, <->] (0.335\textwidth,0.135\textwidth) -- (0.335\textwidth,0.165\textwidth);
	\node at (0.385\textwidth,0.0675\textwidth) {$\SI{22.5}{\milli \meter}$};
	\node at (0.375\textwidth,0.15\textwidth) {$\SI{5}{\milli \meter}$};
	\draw [draw=gray, dashed, line width=0.2mm] (0.3\textwidth,-0.02\textwidth) -- (0.3\textwidth, 0.35\textwidth);
\end{tikzpicture}
	}
	\caption{Specimen with a hole in the middle and phased array on top}
	\label{specimen_circle}
\end{figure}
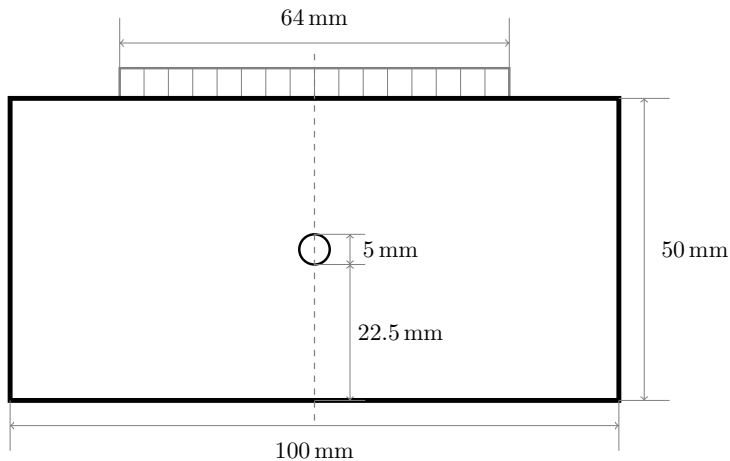

\subsection{Computation of the initial and idealized intermediate gradients}

For a better understanding of the following inversion results, the initial gradients and the gradients at two artificial idealized intermediate states are computed and discussed. Initially, $\gamma = 1$ is set for the whole domain. \figref{gradient_1} shows the gradients for the three parametrizations $\rho$-, $c$- and $\rho c$-scaling. The gradient is computed using the sensitivity kernels of the first time-reversed solution, where the adjoint source term is the difference of signals from the assumed, i.e. still undisturbed model and the measurement at the reference model with void. We then choose $\gamma = 0.6$ in the void as an intermediate state and afterwards decrease $\gamma$ to 0.2 for an even higher contrast. In the healthy domain $\gamma$ remains untouched. We $L^2$ project these piecewise constant scaling functions onto the finite element basis to obtain the corresponding coefficients $\hat{\gamma}_i$ that we compute the adjoint gradient with. \appref{App:gradients} shows the computed gradients for both idealized intermediate states.

For the homogeneous initial model, the gradients in all three formulations are sensitive to a change of the scaling function $\gamma$ within the void region. The size and shape of the defect are properly reproduced. The gradients of the first idealized intermediate state are shown in \figref{gradient_0p6}. The $\rho$-scaling results in a gradient that continues to reproduce the void at the right location and in the right shape. 
Conversely, $c$-scaling causes the gradient within the defect to point in the opposite direction. Instead of further decreasing $\gamma$ in an inversion, this results in an increase of $\gamma$ within the void. This unwanted behavior is caused by the difference in wave speed between the void and the undamaged material. Waves transmitted through the defect experience strong phase shifts, resulting in a distortion of the gradient. In an inversion, transmitted waves are mismatched to unrelated wave packages from multiple reflections. The optimization gets stuck in a physical meaningless local minimum. This undesirable behavior dominates the information contained in the reflected waves.
\begin{figure}[H]
	\centering
	\begin{subfigure}[t]{0.3\textwidth}
		\centering
		\includegraphics[width=\textwidth]{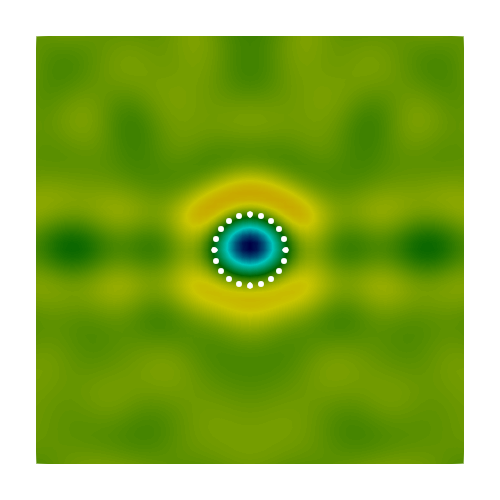}
		\caption{$\rho$-scaling}
	\end{subfigure}
	%\hfill
	\begin{subfigure}[t]{0.3\textwidth}
		\centering
		\includegraphics[width=\textwidth]{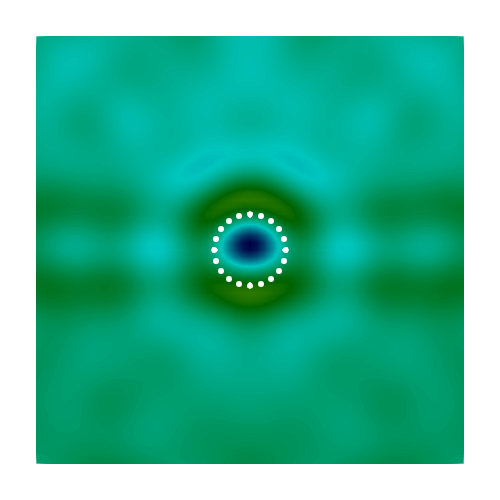}
		\caption{$c$-scaling}
	\end{subfigure}
	%\hfill
	\begin{subfigure}[t]{0.3\textwidth}
		\centering
		\includegraphics[width=\textwidth]{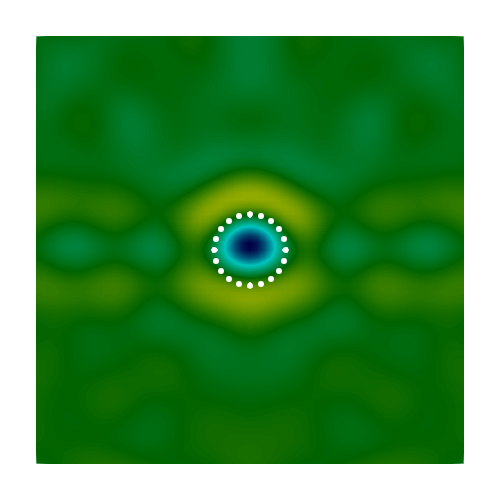}
		\caption{$\rho c$-scaling}
	\end{subfigure}
	%\hfill
	\begin{subfigure}[t]{0.07\textwidth}
		\centering
		\includegraphics{tikz/Colorbar_-1to1.tikz}
	\end{subfigure}
	%hfill
	\caption{Initial gradients with $\gamma = 1$ everywhere}
	\label{gradient_1}
\end{figure}
For $\rho c$-scaling, the reflections are stronger because the impedance is the product of the scaled density and wave speed. The reflection information still pushes the gradient to point to the right direction within the void. Nevertheless, the size is greatly underestimated. For an even higher contrast $\gamma = 0.2$ in the void, see \figref{gradient_0p2}, the gradient continues to point in the right direction for the $\rho$-scaling. As explained, the introduced phase shifts lead to a distortion of the gradient using $c$-scaling. In this configuration, also the gradient of $\rho c$-scaling starts to point in the wrong direction, as the error caused by phase shifts becomes dominant.

\subsection{Inversion results}

\figref{objective_circle} shows the development of the objective function for all four formulations normalized with respect to the value of the initial model. It can be seen that the FWI converges by far the fastest with $\rho$-scaling. After only $5$ iterations, the cost function falls below $\SI{10}{\percent}$ of its initial value. \figref{circleRho} shows the corresponding inverted scaling function $\gamma$ for the iterations $5$, $10$ and $25$.

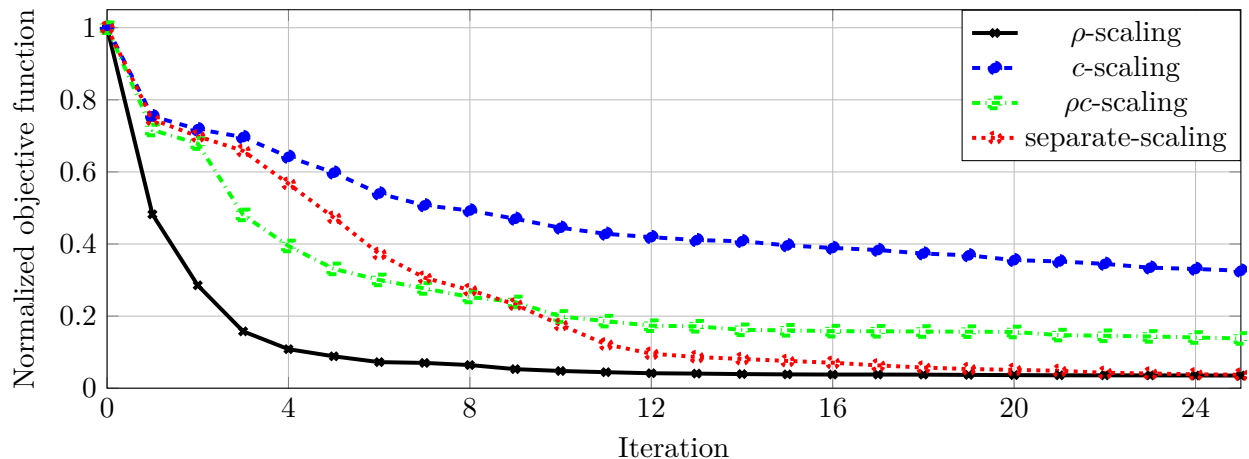
\begin{figure}[H]
	\centering
	\begin{tikzpicture}
	\begin{axis}[
		xmin = 0, xmax = 25,
		ymin = 0, ymax = 1.05,
		xtick distance = 4,
		ytick distance = 0.2,
		grid=both,
		width = \textwidth,
		height = 0.4\textwidth,
		xlabel = {Iteration},
		ylabel = {Normalized objective function},
		legend style={at={(1,1)}, anchor=north east}]
		
		\addplot[black, mark = x, line width=1.5pt] file[] {tikz/circle_rho_costs.dat};
		\addplot[blue, dashed, mark = *, line width=1.5pt] file[] {tikz/circle_c_costs.dat};
		\addplot[green, dash dot, mark = square, line width=1.5pt] file[] {tikz/circle_rhoc_costs.dat};
		\addplot[red, dotted, mark = o, line width=1.5pt] file[] {tikz/circle_2P_costs.dat};
		
		\legend{$\rho$-scaling, $c$-scaling, $\rho c$-scaling, separate-scaling}
	\end{axis}
\end{tikzpicture}
	\vspace{-0.75cm}
	\caption{Development of the normalized objective function for the reconstruction of a circular void}
	\label{objective_circle}
\end{figure}

\begin{figure}[H]
	\centering
	\begin{subfigure}[t]{0.3\textwidth}
		\centering
		\includegraphics[width=\textwidth]{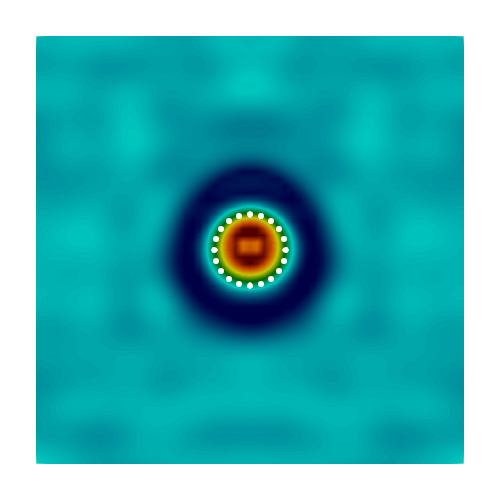}
		\vspace{-0.75cm}
		\caption{Iteration $5$}
	\end{subfigure}
	%\hfill
	\begin{subfigure}[t]{0.3\textwidth}
		\centering
		\includegraphics[width=\textwidth]{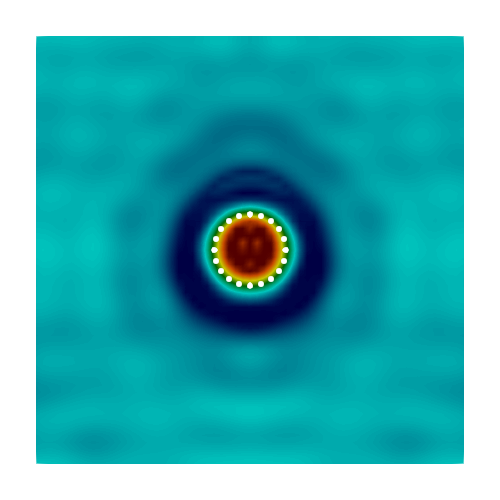}
		\vspace{-0.75cm}
		\caption{Iteration $10$}
	\end{subfigure}
	%\hfill
	\begin{subfigure}[t]{0.3\textwidth}
		\centering
		\includegraphics[width=\textwidth]{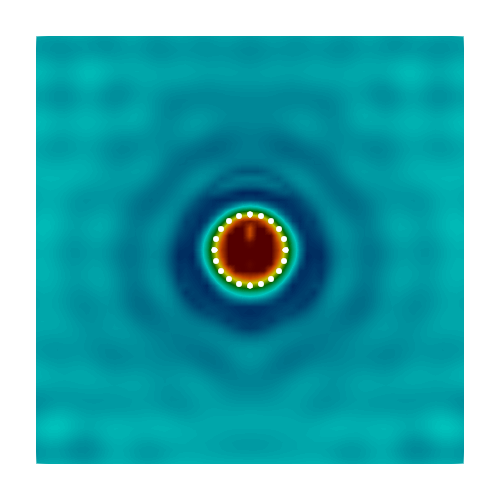}
		\vspace{-0.75cm}
		\caption{Iteration $25$}
	\end{subfigure}
	%\hfill
	\begin{subfigure}[t]{0.07\textwidth}
		\centering
		\includegraphics{tikz/Colorbar_0to1p2.tikz}
	\end{subfigure}
	%hfill
	\caption{$\rho$-scaling: reconstructed scaling function $\gamma$ for a circular void}
	\label{circleRho}
\end{figure}

For $\rho$-scaling, the position and shape of the circular void is correctly reconstructed already after $5$ iterations. However, the size is slightly underestimated. As the number of iterations increases, the outline of the circular void is reconstructed almost perfectly. Additionally, the surrounding artifacts decrease.

The $c$-scaling approach shows the worst convergence of all parametrizations. After $25$ iterations the objective function is still at $\SI{32.6}{\percent}$ of its initial value. The intermediate states of $\gamma$ shown in \figref{circleC} indicate that the inversion begins to reconstruct the circular void at the right position with a circular shape. As expected from the preliminary computations of gradients at idealized intermediate states, the shape is distorted for a higher difference in the wave speed at later iterations. This distortion is caused by the introduced phase shifts of the transmitted waves and the occurrence of unwanted oscillations at material interfaces with high contrast in the wave speed. The void is reconstructed in such a way that the reflection behavior is still well approximated, while the area of the void is as small as possible to avoid phase shifts of the transmitted waves. The optimization converges to a local minimum where the model has a horizontal 'slit' with a small extension in the vertical direction. Strong artifacts appear in the area sourrounding the defect. 

The $\rho c$-scaling results in a normalized objective functions of $\SI{13.8}{\percent}$ after $25$ iterations. Considering the scaling function $\gamma$ after $5$, $10$ and $25$ iterations (\figref{circleRhoC}), the void is reconstructed at the right location. The circular shape is also restored. However, due to the phase shift discussed in \secref{preliminaryForward}, the inversion significantly underestimates the size of the circular void. There are only minor artifacts around the defect.

\begin{figure}[H]
	\centering
	\begin{subfigure}[t]{0.3\textwidth}
		\centering
		\includegraphics[width=\textwidth]{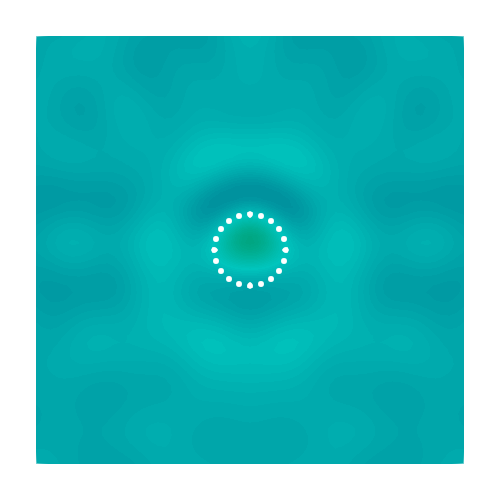}
		\caption{Iteration $5$}
	\end{subfigure}
	%\hfill
	\begin{subfigure}[t]{0.3\textwidth}
		\centering
		\includegraphics[width=\textwidth]{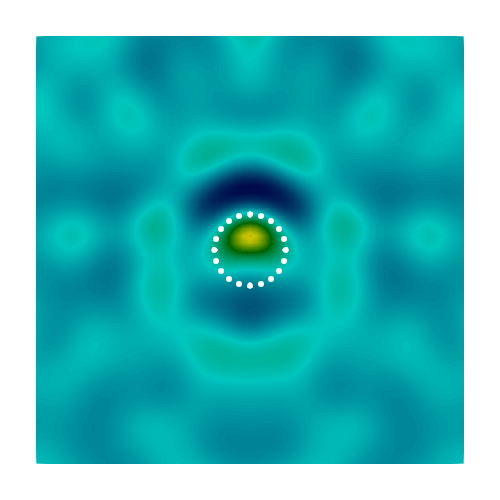}
		\caption{Iteration $10$}
	\end{subfigure}
	%\hfill
	\begin{subfigure}[t]{0.3\textwidth}
		\centering
		\includegraphics[width=\textwidth]{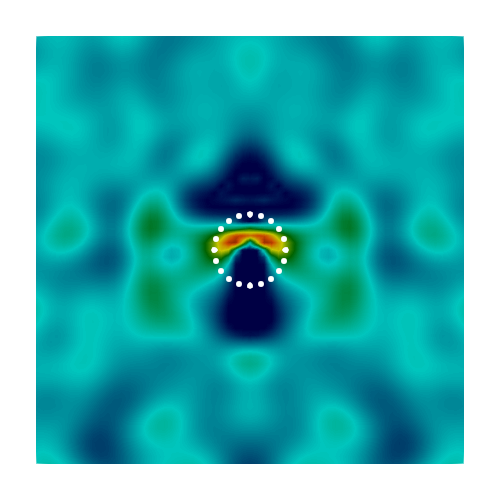}
		\caption{Iteration $25$}
	\end{subfigure}
	%\hfill
	\begin{subfigure}[t]{0.07\textwidth}
		\centering		
		\includegraphics{tikz/Colorbar_0to1p2.tikz}
	\end{subfigure}
	%hfill
	\caption{$c$-scaling: reconstructed scaling function $\gamma$ for a circular void}
	\label{circleC}
\end{figure}

\begin{figure}[H]
	\centering
	\begin{subfigure}[t]{0.3\textwidth}
		\centering
		\includegraphics[width=\textwidth]{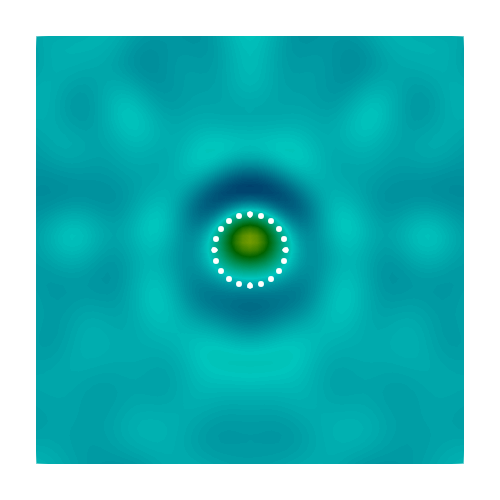}
		\caption{Iteration $5$}
	\end{subfigure}
	%\hfill
	\begin{subfigure}[t]{0.3\textwidth}
		\centering
		\includegraphics[width=\textwidth]{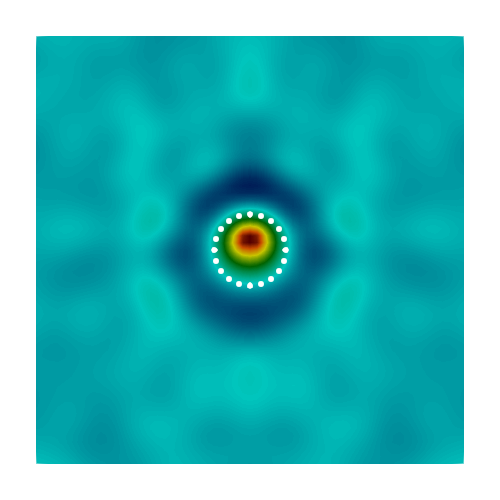}
		\caption{Iteration $10$}
	\end{subfigure}
	%\hfill
	\begin{subfigure}[t]{0.3\textwidth}
		\centering
		\includegraphics[width=\textwidth]{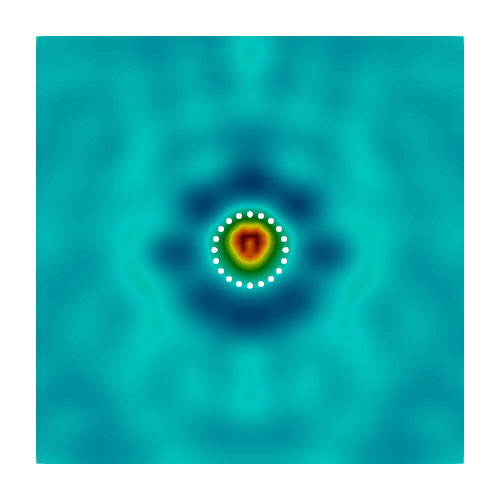}
		\caption{Iteration $25$}
	\end{subfigure}
	%\hfill
	\begin{subfigure}[t]{0.07\textwidth}
		\centering
		\includegraphics{tikz/Colorbar_0to1p2.tikz}
	\end{subfigure}
	%hfill
	\caption{$\rho c$-scaling: reconstructed scaling function $\gamma$ for a circular void}
	\label{circleRhoC}
\end{figure}

The development of the objective function for the two-parameter FWI approach closely follows the $c$-scaling of mono-parameter FWI in the first $3$ iterations. After that, the normalized objective function decreases much faster, leading to a final value that is very close to the $\rho$-scaling. \figref{circleSeparate} shows the inverted scaling functions of density $\gamma_\rho$ and of wave speed $\gamma_c$ at the iterations $5$, $10$ and~$25$. Here we observe that the reconstruction changes both scaling functions evenly at the beginning. After the first $5$ iterations, the inversion focuses mainly on updating $\gamma_\rho$. The position, shape and size of the defect are accurately reproduced by $\gamma_\rho$.

\begin{figure}[H]
	\centering
	\begin{subfigure}[t]{0.3\textwidth}
		\centering
		\includegraphics[width=\textwidth]{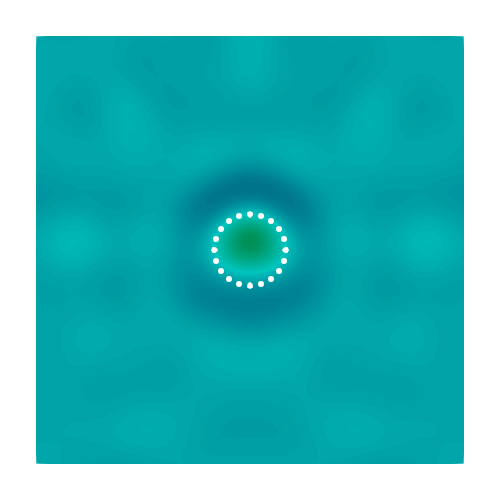}
		\caption{Iteration $5$ -- $\gamma_\rho$}
	\end{subfigure}
	%\hfill
	\begin{subfigure}[t]{0.3\textwidth}
		\centering
		\includegraphics[width=\textwidth]{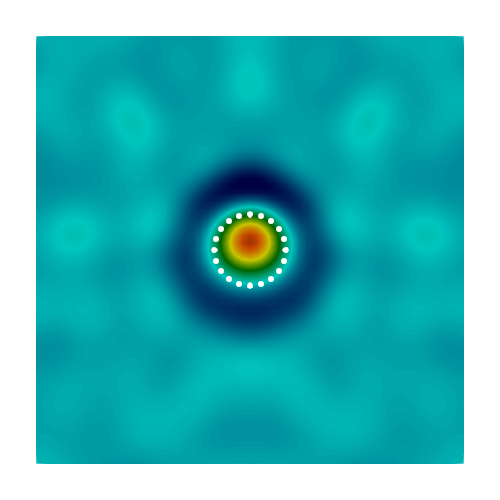}
		\caption{Iteration $10$ -- $\gamma_\rho$}
	\end{subfigure}
	%\hfill
	\begin{subfigure}[t]{0.3\textwidth}
		\centering
		\includegraphics[width=\textwidth]{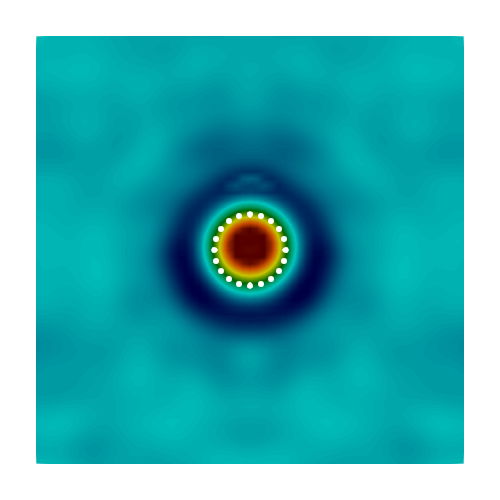}
		\caption{Iteration $25$ -- $\gamma_\rho$}
	\end{subfigure}
	%\hfill
	\begin{subfigure}[t]{0.07\textwidth}
		\centering
		\includegraphics{tikz/Colorbar_0to1p2.tikz}
	\end{subfigure}
	%hfill
	\begin{subfigure}[t]{0.3\textwidth}
		\centering
		\includegraphics[width=\textwidth]{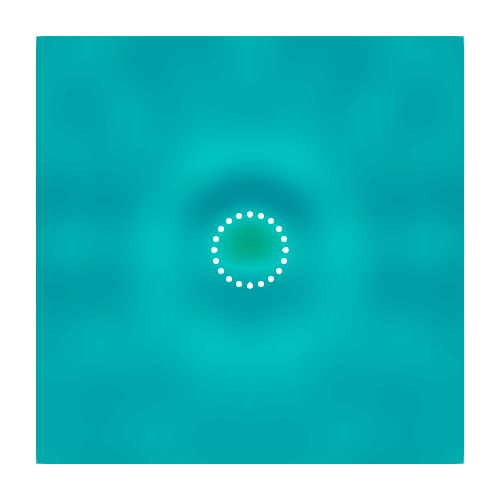}
		\caption{Iteration $5$ -- $\gamma_c$}
	\end{subfigure}
	%\hfill
	\begin{subfigure}[t]{0.3\textwidth}
		\centering
		\includegraphics[width=\textwidth]{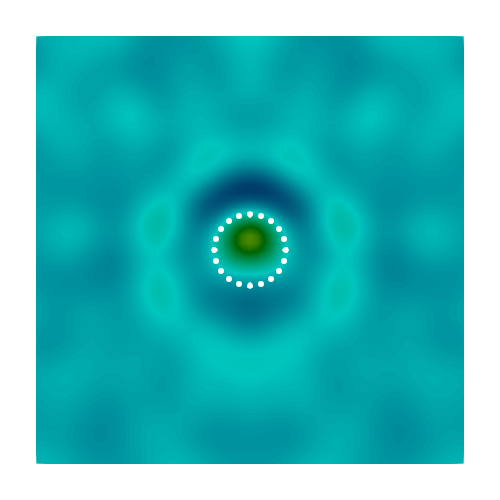}
		\caption{Iteration $10$ -- $\gamma_c$}
	\end{subfigure}
	%\hfill
	\begin{subfigure}[t]{0.3\textwidth}
		\centering
		\includegraphics[width=\textwidth]{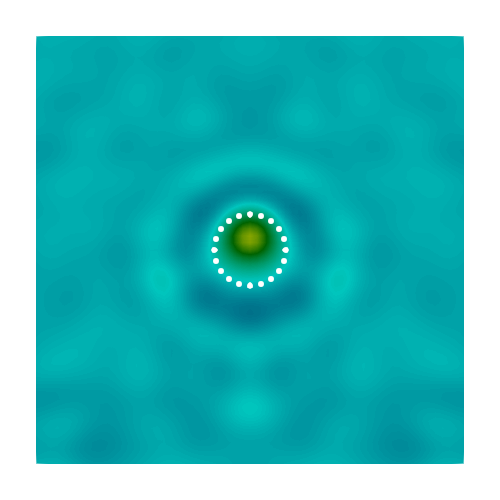}
		\caption{Iteration $25$ -- $\gamma_c$}
	\end{subfigure}
	%\hfill
	\begin{subfigure}[t]{0.07\textwidth}
		\centering
		\includegraphics{tikz/Colorbar_0to1p2.tikz}
	\end{subfigure}
	%hfill
	\caption{separate-scaling: reconstructed scaling functions $\gamma_\rho$ and $\gamma_c$ for a circular void}
	\label{circleSeparate}
\end{figure}

The results show that modeling voids in the inverse problem with a material parameter scaling only the density leads to the best results. The $\rho$-scaling far outperforms the other mono-parameter FWI formulations and also shows a better convergence behavior than the two-parameter FWI. The preliminary investigations on the forward problem and gradients show that scaling the wave speed at high material contrasts leads to oscillations in the physical domain for the forward solution and distorted gradients due to undesired phase shifts and mismatching of wave packages. This behavior leads to a convergence to physical meaningless local minima in the inversion. The results of the two-parameter inversion strengthen these observations. Instead of reconstructing both material parameters, the inversion focuses on the density while ignoring the wave speed. Since the $\rho$-scaling has the best performance of all presented variants, we only combine this approach with the finite cell method (FCM) for the reconstruction of voids in arbitrary complex domains in the following section.
}

\section{Voids in an immersed setting}
{
\label{sec:ellipticalvoid}

\subsection{Finite cell method}
{
\label{sec:FCM}

In the previous sections the variation of material parameters was used to model structural details like areas of lower densities or void regions with (close to) zero density. This approach is conceptually similar to immersed boundary methods (IBM), where a known structure is embedded in a larger, simply shaped 'fictitious domain', and an indicator function $\alpha$ discriminates between the fictitious and physical part of the computation (see \figref{embeddedDomain}). While ‘fictitious domain methods’ were suggested already in the 1960ies for a numerical approximation of boundary value problems \cite{Saulev1963}, they have become very popular again during the last decade, in particular as they offer attractive opportunities for closely integrating geometric modeling and numerical analysis. Many variants of immersed boundary methods have been proposed. Among these are the finite cell method (FCM \cite{Duester2008}), its IGA-based variant immersogeometric analysis \cite{Schillinger2012}, CutFEM \cite{Burman2015}, IBRA \cite{Breitenberger2015}, the aggregated finite element method \cite{Badia2018}, the cgFEM \cite{Nadal2013} or the shifted boundary method \cite{Main2018_1, Main2018_2}. CutFEM \cite{Sticko2020} and the finite cell method have also been extended to scalar and elastic wave equations. The spectral cell method \cite{Duczek2014, Joulaian2014, Nicoli2022} concentrates on an efficient combination of immersed boundary approaches and explicit time integration for wave equations. We will exemplify the application of full waveform inversion for immersed boundary approaches concerning in particular the finite cell method.

\begin{figure}[H]
	\centering
	\resizebox{0.8\textwidth}{!}{
		\input{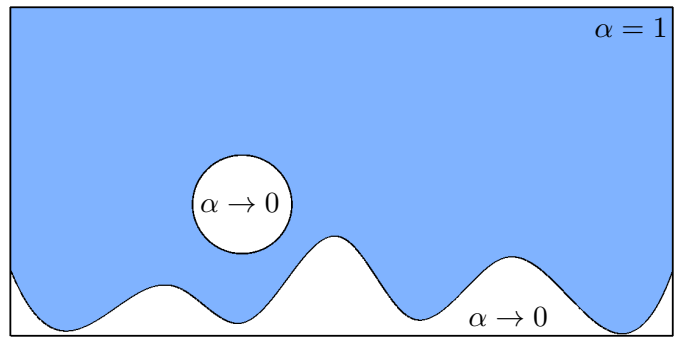}
	}
	\vspace{-1cm}
	\caption{A priori known geometry incorporated by the indicator function $\alpha$. Blue: Physical domain, white: Fictitious domain}
	\label{embeddedDomain}
\end{figure}

The major asset of FCM like that of other immersed boundary methods is obvious: The physical domain of computation needs not be meshed into finite elements, which is, in particular in cases of complex three-dimensional geometries still a major issue for many applications. Instead of a mesh of boundary-conforming elements the simply shaped fictitious domain is divided into a Cartesian grid of uniform cells. The such defined grid is then used as a 'mesh' of cells, where each cell is treated like a finite element. The central advantage of saving the effort to generate a mesh yet comes at a significant cost, and several critical questions need to be answered to obtain a feasible numerical method. First of all, cells cut by the boundary of the physical domain need special treatment during the computation of element mass and stiffness matrices and load vectors, as their integrand is discontinuous. Space-tree based numerical integration is a simple and robust possibility~\cite{Duester2008}. Construction of specific integration formulae \cite{Joulaian2016}, the generation of local integration meshes \cite{Fries2016} or a local modification of the boundary conditions \cite{Main2018_1} are others. A further issue is the imposition of boundary conditions along segments which are not matching with edges of grid cells. There, Neumann and Dirichlet boundary conditions need to be applied in a weak sense. Simple penalty methods and Nitsche-type formulations \cite{Larsson2022} can successfully be applied. Further consideration needs to be given to cells which have only a small cut part interior to the physical domain. These may result in stability problems of the discretized PDE operator and conditioning problems of the system matrix that can be stabilized, e.g., by coosing a small non-zero value for alpha inside the fictitious domain or by adding ghost penalties across the element interfaces \cite{Burman2010}. The resulting systems can be preconditioned using, e.g.,~\cite{dePrenter2017, Larsson2022, dePrenter2022}}.

We now formulate the strong form of the scalar wave equation in the immersed boundary setting sketched in \figref{embeddedDomain}:
\begin{equation}
	\alpha(\tensor{x}) \rho_0 \ddot{u}(\tensor{x}, t) - \nabla \cdot \left( \alpha(\tensor{x}) \rho_0 c_0^2 \nabla u(\tensor{x}, t) \right) = f(\tensor{x}, t)
\end{equation}
with the known indicator function $\alpha$. It takes the value $1$ inside the physical domain and $0$ (or a small stability preserving positive value, see \cite{Duester2008}) in the fictitious part of the domain including a priori known voids. The combination of this immersed formulation with our $\rho$-scaling results in 
 
\begin{equation}
	\alpha(\tensor{x}) \gamma(\tensor{x}) \rho_0 \ddot{u}(\tensor{x}, t) - \nabla \cdot \left( \alpha(\tensor{x}) \gamma(\tensor{x}) \rho_0 c_0^2 \nabla u(\tensor{x}, t) \right) = f(\tensor{x}, t)\text{.}
\end{equation}
Applying now full waveform inversion (FWI), unknown voids are iteratively characterized by reconstructing the scaling function $\gamma$. In summary, the a priori defined indicator function $\alpha$ is superposed with the iteratively updated scaling function $\gamma$, see \figref{embeddedDomain_withFlaw}.

\begin{figure}[H]
	\centering
	\resizebox{0.8\textwidth}{!}{
		\input{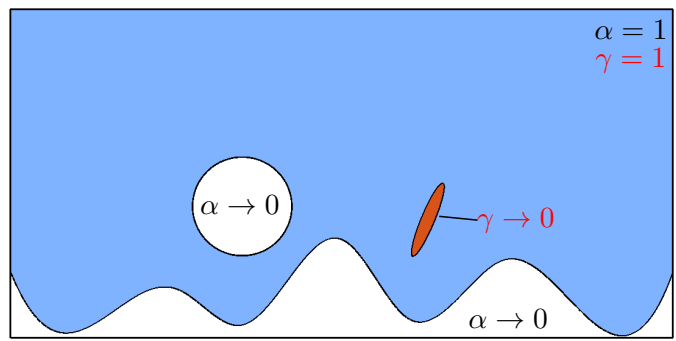}
	}
	\vspace{-1cm}
	\caption{A priori known geometry incorporated by the indicator function $\alpha$ with an unknown void reconstructed by the scaling function $\gamma$}
	\label{embeddedDomain_withFlaw}
\end{figure}

\subsection{Configuration}

In the following inversion example an elliptical void is identified in an immersed boundary setup. \figref{embeddedDomain_withFlaw} shows the extended domain. The inversion is performed with the $\rho$-scaling formulation in combination with the FCM. Again, the synthetic reference data is computed using a boundary-conforming unstructured mesh of $50\,192$ linear quadrilateral elements generated using Gmsh. For the inversion model, an extended domain of size $\SI{100}{\milli \meter} \times \SI{50}{\milli \meter}$ is used applying the FCM. The density and wave speed of the healthy intact material are $\rho_0 = \SI{2700}{\kilo / \meter^3}$ and $c_0 = \SI{6000}{\meter / \second}$. The extended domain is meshed with uniform linear quadrilateral elements of size $h_e = \SI{0.5}{\milli \meter}$, resulting in $20\,000$ elements. The circular void at position $x = \SI{35}{\milli \meter}$ and $y = \SI{20}{\milli \meter}$ with the radius $r = \SI{7.5}{\milli \meter}$ and the bottom surface are known a priori and are incorporated by setting the FCM indicator function $\alpha = \SI{1e-3}{}$ within the void regions and to $1$ everywhere else. The bottom surface is discretized by cubic splines defined by the interpolation points 
$P_1 = (0, \SI{10}{\milli \meter})$, 
$P_2 = (\SI{10}{\milli \meter}, \SI{1}{\milli \meter})$, 
$P_3 = (\SI{25}{\milli \meter}, \SI{7.5}{\milli \meter})$, 
$P_4 = (\SI{35}{\milli \meter}, \SI{2}{\milli \meter})$, 
$P_5 = (\SI{50}{\milli \meter}, \SI{15}{\milli \meter})$, 
$P_6 = (\SI{60}{\milli \meter}, \SI{3}{\milli \meter})$, 
$P_7 = (\SI{75}{\milli \meter}, \SI{12}{\milli \meter})$, 
$P_8 = (\SI{90}{\milli \meter}, \SI{1}{\milli \meter})$ and
$P_9 = (\SI{100}{\milli \meter}, \SI{10}{\milli \meter})$.

A quadtree of partitioning depth $5$ is used for the integration of the cut elements. For both the reference and inversion simulations, the simulation time is $T = \SI{6.0e-5}{\second}$, the time step size is $\Delta t = \SI{5.0e-9}{\second}$ and second-order central differences are applied for time integration. A 2-cycle sine burst with a central frequency at $f = \SI{500}{\kilo \hertz}$ is used for excitation. An array of 65 transducers with a pitch of 1 mm is placed at the center of the top surface. The unknown ellipse at position $x = \SI{63}{\milli \meter}$ and $y = \SI{18}{\milli \meter}$ has the width $a = \SI{6}{\milli \meter}$ and the height $b = \SI{1}{\milli \meter}$ and is rotated by $\SI{67.5}{\degree}$. It is reconstructed using the $\rho$-scaling of the FWI. The a priori known void regions, i.e. the circle and the background, are excluded from the inversion domain. In the remaining domain, the scaling function is clipped between $0.0$ and $1.2$ during the inversion. For $\rho$-scaling $25$~L-BFGS-B iterations are executed.

\subsection{Inversion results}

\figref{objective_ellipse} shows the evolution of the objective function normalized with respect to its initial value for the FWI using the $\rho$-scaling approach in combination with the FCM. The objective function steadily decreases, falling below $\SI{10}{\percent}$ of its initial value after only $10$ iterations. After the full $25$ iterations the normalized objective is $\SI{2.3}{\percent}$. 

After $5$ iterations the location, shape and size of the ellipse are already reproduced almost perfectly. However, considerable artifacts are observed, especially in the area around the defect. While the shape of the ellipse does not change significantly as the number of iterations increases, the artifacts decrease successively, leading to a very small disturbance of the final reconstruction. The incorporation of the a priori known geometric features using the FCM does not affect the quality of the inversion with $\rho$-scaling in any way. The fictitious domain is excluded from the inversion domain since it is known a priori. The superposition of the indicator function $\alpha$ and scaling function $\gamma$ does not degrade the quality of the forward simulations in the inversion framework. The FWI provides a smoothed approximation of the elliptic void by the scaling function $\gamma$ and reaches a value close to $0$ in the defect, while the a priori known features are described by the discontinuous indicator function~$\alpha$. 

\begin{figure}[H]
	\centering
	\begin{tikzpicture}
	\begin{axis}[
		xmin = 0, xmax = 25,
		ymin = 0, ymax = 1.05,
		xtick distance = 4,
		ytick distance = 0.2,
		grid=both,
		width = \textwidth,
		height = 0.4\textwidth,
		xlabel = {Iteration},
		ylabel = {Normalized objective function},
		legend style={at={(1,1)}, anchor=north east}]		
		\addplot[black, mark = x, line width=1.5pt] file[] {tikz/ellipse_rho_costs.dat};
		\legend{$\rho$-scaling}
	\end{axis}
\end{tikzpicture}
	\caption{Development of the normalized objective function for the reconstruction of an elliptical void}
	\label{objective_ellipse}
\end{figure}
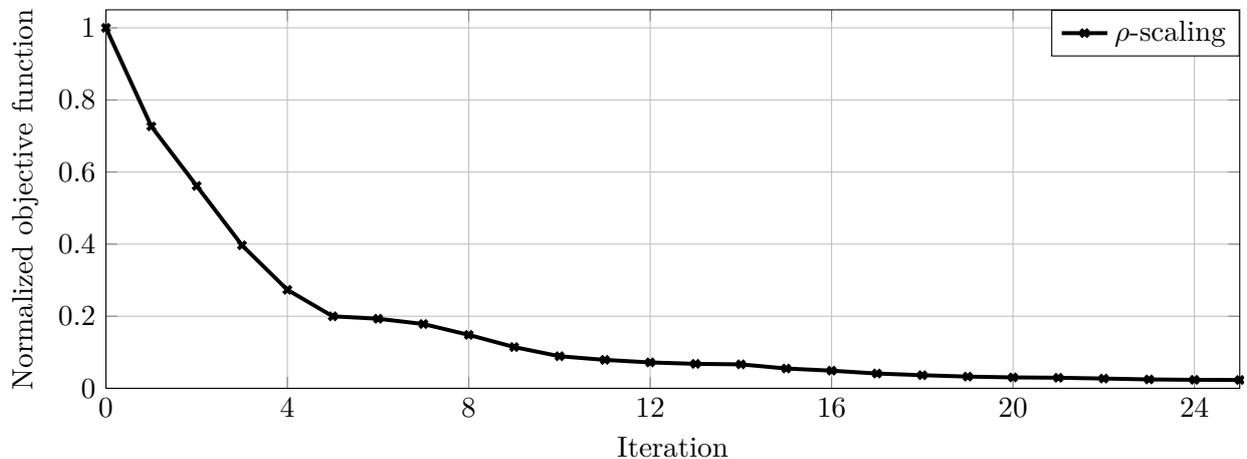

\section{Conclusion}
{ 
\label{sec:conclusion}
	
In this contribution, the forward and inverse scalar wave equation in domains with voids is considered. Three different parametrizations with a single material parameter and one variant with two parameters are presented and compared. Initial investigations show that modeling voids by adjusting only the wave speed leads to erroneous behavior in the forward simulation. Unwanted oscillations are induced back into the physical domain, after a wave hits the interface to the void. In contrast, density scaling proves to be very suitable for forward simulation. The computation of the gradients of the inverse problem again indicates difficulties in modeling voids by downscaling the wave speed. The resulting phase shifts and the associated wave package mismatches lead to a distortion of the gradients. Density scaling again leads to superior behavior over all other approaches. The gradient still excellently detects the defect even at high contrasts in the material. In agreement with the previous observations, the density scaling approach also outperforms the other approaches when reconstructing a circular void in a plate. The two-parameter formulation confirms these findings, as after only a few iterations the wave speed remains nearly unchanged while the density continues to be updated. 

\begin{figure}[H]
	\centering
	\begin{subfigure}[t]{0.55\textwidth}
		\centering
		\includegraphics[width=\textwidth]{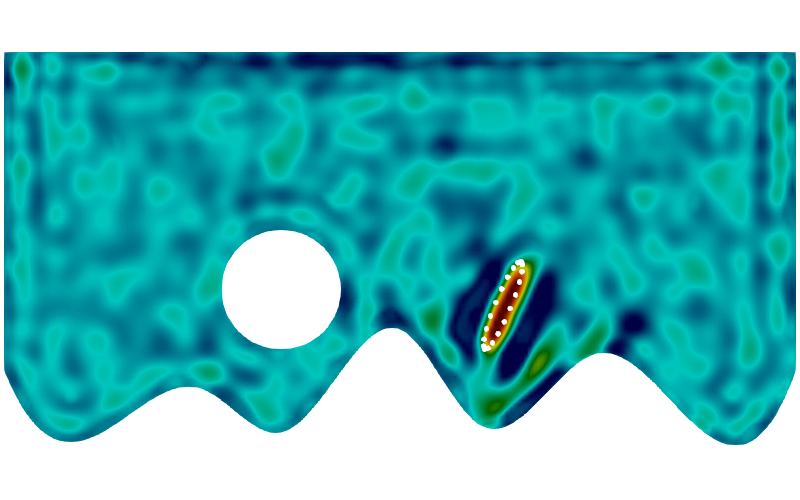}
		\caption{Iteration $5$}
	\end{subfigure}
	\hspace{1.0cm}
	\hspace{-1.0cm}
	\begin{subfigure}[t]{0.07\textwidth}
		\centering
		\vspace{-0.8cm}
		\includegraphics{tikz/Colorbar_0to1p2.tikz}
	\end{subfigure}
	\begin{subfigure}[t]{0.55\textwidth}
		\centering
		\includegraphics[width=\textwidth]{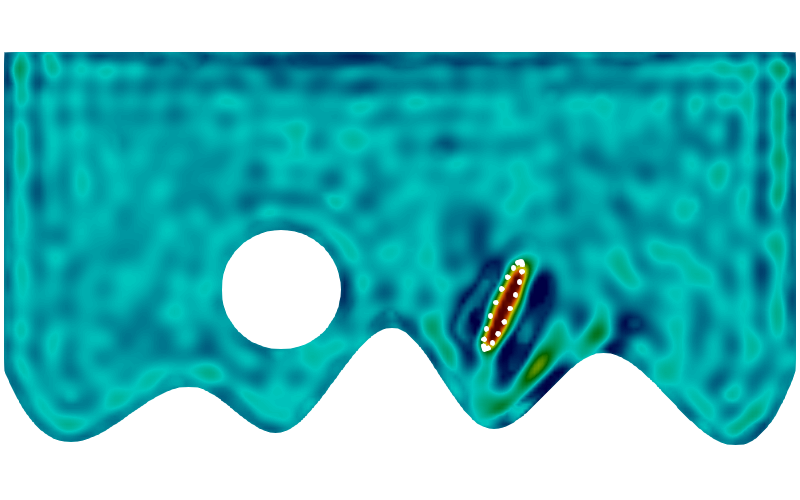}
		\caption{Iteration $10$}
	\end{subfigure}
	\hspace{1.0cm}
	\hspace{-1.0cm}
	\begin{subfigure}[t]{0.07\textwidth}
		\centering
		\vspace{-0.8cm}
		\includegraphics{tikz/Colorbar_0to1p2.tikz}
	\end{subfigure}
	\begin{subfigure}[t]{0.55\textwidth}
		\centering
		\input{tikz/background_rho_it25_withZoom.tex}
		\caption{Iteration $25$}
	\end{subfigure}
	\hspace{2.0cm}
	\begin{subfigure}[t]{0.07\textwidth}
		\centering
		\vspace{-1.2cm}
		\includegraphics{tikz/Colorbar_0to1p2.tikz}
	\end{subfigure}
	\hspace{-1.2cm}
	%hfill
	\caption{$\rho$-scaling: reconstructed scaling function $\gamma$ for an elliptical void}
	\label{gamma_ellipse}
\end{figure}
	
The density scaling approach is combined with the finite cell method, immersing the domain of computation in a larger, simply shaped fictitious domain. While a priori known geometric features are accounted for by a discontinuous indicator function $\alpha$, the smooth scaling function $\gamma$, computed iteratively in the full waveform inversion, recovers unknown defects. Within the defect region, $\gamma$ tends to a value close to $0$, while only minor artifacts occur elsewhere. For the inversion of an elliptic void, the combination of density scaled full waveform inversion and the finite cell method proves to be a powerful tool. After only a few iterations, the defect is almost perfectly reproduced. Given this proof of concept for the scalar wave equation, the authors plan to extend the approach to the elastic wave equation, to 3-dimensional problems and apply it to experimental data in the future. 

}

}

\appendix
\section*{Appendices}
\section{1D forward problem}
\label{App:1D}
{
	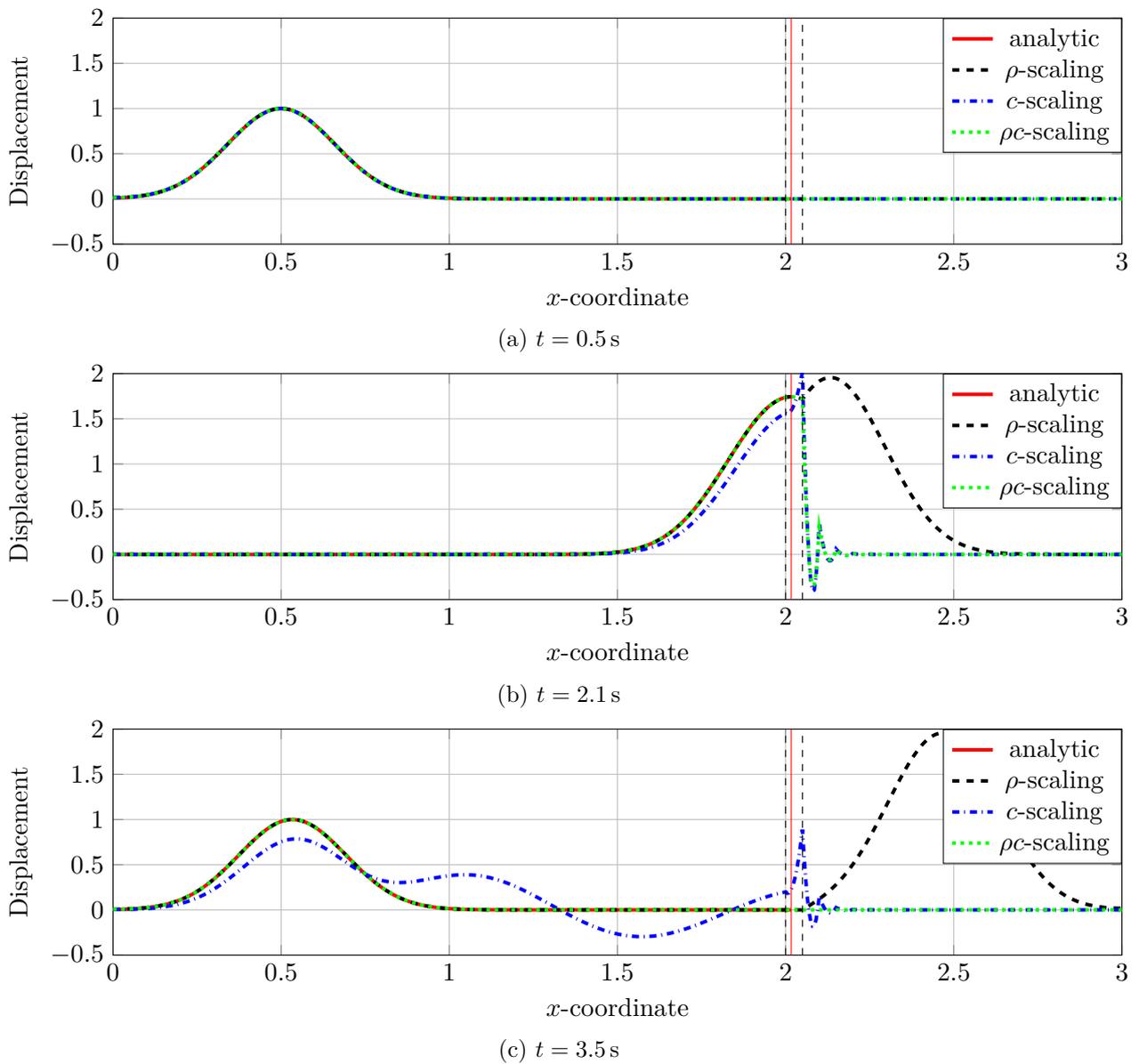
\begin{figure}[H]
		\centering
		\begin{subfigure}[b]{1.0\textwidth}
			\centering
			\begin{tikzpicture}
	\begin{axis}[
		xmin = 0, xmax = 3,
		ymin = -0.5, ymax = 2.0,
		xtick distance = 0.5,
		ytick distance = 0.5,
		grid=both,
		width = \textwidth,
		height = 0.3\textwidth,
		xlabel = {$x$-coordinate},
		ylabel = {Displacement},
		legend style={at={(1,1)}, anchor=north east}]
		
		\addplot[red, line width=1.5pt] file[] {tikz/data_1Dforward/1Dforward_analytic_t1.dat};
		\addplot[black, dashed, line width=1.5pt] file[] {tikz/data_1Dforward/1Dforward_p2_v0_t1.dat};
		\addplot[blue, dash dot, line width=1.5pt] file[] {tikz/data_1Dforward/1Dforward_p2_v1_t1.dat};
		\addplot[green, dotted, line width=1.5pt] file[] {tikz/data_1Dforward/1Dforward_p2_v2_t1.dat};
		\addplot[black, dashed] file[] {tikz/data_1Dforward/elBoundary1.dat};
		\addplot[black, dashed] file[] {tikz/data_1Dforward/elBoundary2.dat};
		\addplot[red] file[] {tikz/data_1Dforward/interface.dat};
		
		\legend{analytic, $\rho$-scaling, $c$-scaling, $\rho c$-scaling}
	\end{axis}
\end{tikzpicture}
			\vspace{-0.5cm}
			\caption{$t = \SI{0.5}{\second}$}
		\end{subfigure}
		%\hfill
		\begin{subfigure}[b]{1.0\textwidth}
			\centering
			\begin{tikzpicture}
	\begin{axis}[
		xmin = 0, xmax = 3,
		ymin = -0.5, ymax = 2.0,
		xtick distance = 0.5,
		ytick distance = 0.5,
		grid=both,
		width = \textwidth,
		height = 0.3\textwidth,
		xlabel = {$x$-coordinate},
		ylabel = {Displacement},
		legend style={at={(1,1)}, anchor=north east}]
		
		\addplot[red, line width=1.5pt] file[] {tikz/data_1Dforward/1Dforward_analytic_t3.dat};
		\addplot[black, dashed, line width=1.5pt] file[] {tikz/data_1Dforward/1Dforward_p2_v0_t3.dat};
		\addplot[blue, dash dot, line width=1.5pt] file[] {tikz/data_1Dforward/1Dforward_p2_v1_t3.dat};
		\addplot[green, dotted, line width=1.5pt] file[] {tikz/data_1Dforward/1Dforward_p2_v2_t3.dat};
		\addplot[black, dashed] file[] {tikz/data_1Dforward/elBoundary1.dat};
		\addplot[black, dashed] file[] {tikz/data_1Dforward/elBoundary2.dat};
		\addplot[red] file[] {tikz/data_1Dforward/interface.dat};
		
		\legend{analytic, $\rho$-scaling, $c$-scaling, $\rho c$-scaling}
	\end{axis}
\end{tikzpicture}
			\vspace{-0.5cm}
			\caption{$t = \SI{2.1}{\second}$}
		\end{subfigure}
		%\hfill
		\begin{subfigure}[b]{1.0\textwidth}
			\centering
			\begin{tikzpicture}
	\begin{axis}[
		xmin = 0, xmax = 3,
		ymin = -0.5, ymax = 2.0,
		xtick distance = 0.5,
		ytick distance = 0.5,
		grid=both,
		width = \textwidth,
		height = 0.3\textwidth,
		xlabel = {$x$-coordinate},
		ylabel = {Displacement},
		legend style={at={(1,1)}, anchor=north east}]
		
		\addplot[red, line width=1.5pt] file[] {tikz/data_1Dforward/1Dforward_analytic_t4.dat};
		\addplot[black, dashed, line width=1.5pt] file[] {tikz/data_1Dforward/1Dforward_p2_v0_t4.dat};
		\addplot[blue, dash dot, line width=1.5pt] file[] {tikz/data_1Dforward/1Dforward_p2_v1_t4.dat};
		\addplot[green, dotted, line width=1.5pt] file[] {tikz/data_1Dforward/1Dforward_p2_v2_t4.dat};
		\addplot[black, dashed] file[] {tikz/data_1Dforward/elBoundary1.dat};
		\addplot[black, dashed] file[] {tikz/data_1Dforward/elBoundary2.dat};
		\addplot[red] file[] {tikz/data_1Dforward/interface.dat};
		
		\legend{analytic, $\rho$-scaling, $c$-scaling, $\rho c$-scaling}
	\end{axis}
\end{tikzpicture}
			\vspace{-0.5cm}
			\caption{$t = \SI{3.5}{\second}$}
		\end{subfigure}
		%hfill
		\caption{1D interface problem $p = 2$}
		\label{1D_p2}
	\end{figure}
	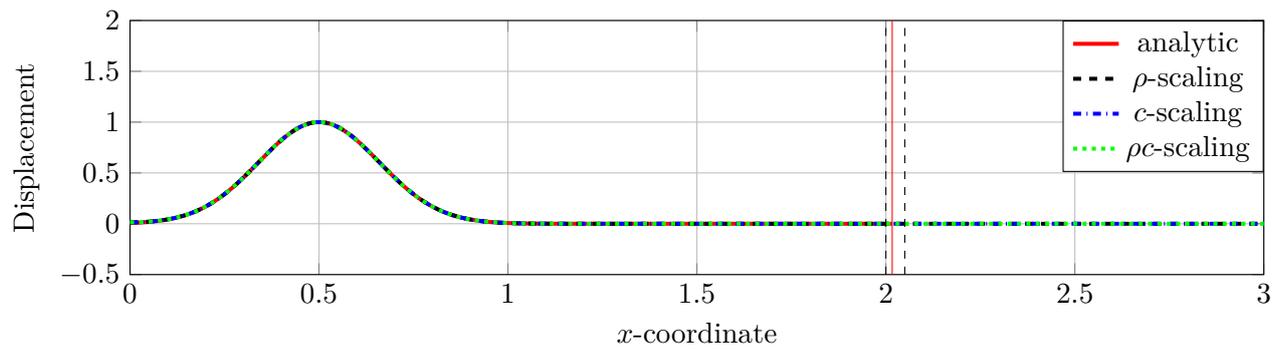
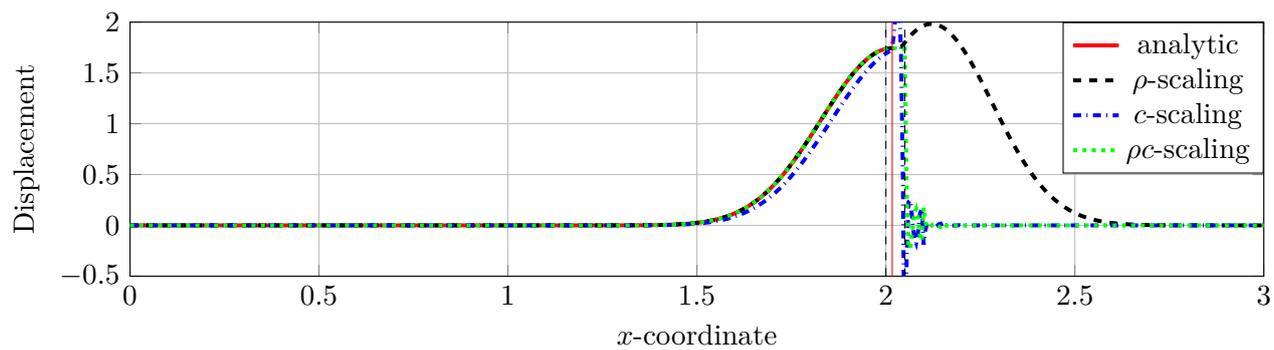
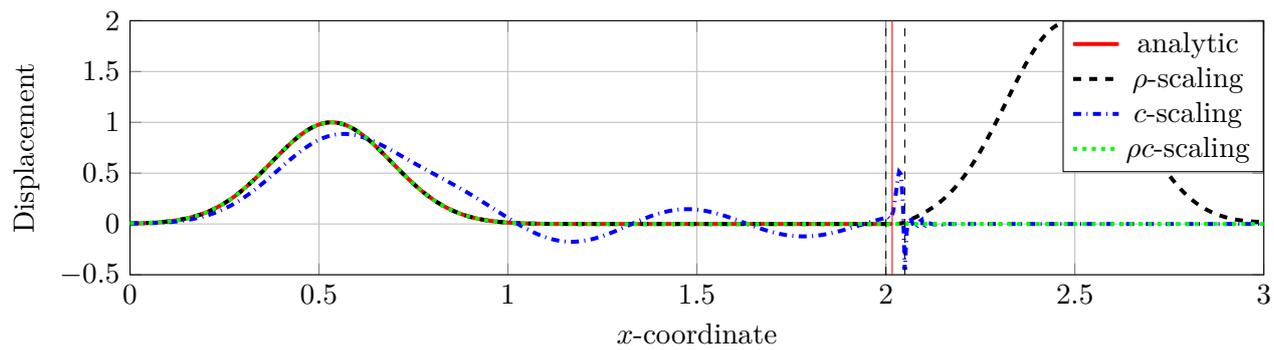
\begin{figure}[H]
		\centering
		\begin{subfigure}[b]{1.0\textwidth}
			\centering
			\begin{tikzpicture}
	\begin{axis}[
		xmin = 0, xmax = 3,
		ymin = -0.5, ymax = 2.0,
		xtick distance = 0.5,
		ytick distance = 0.5,
		grid=both,
		width = \textwidth,
		height = 0.3\textwidth,
		xlabel = {$x$-coordinate},
		ylabel = {Displacement},
		legend style={at={(1,1)}, anchor=north east}]
		
		\addplot[red, line width=1.5pt] file[] {tikz/data_1Dforward/1Dforward_analytic_t1.dat};
		\addplot[black, dashed, line width=1.5pt] file[] {tikz/data_1Dforward/1Dforward_p4_v0_t1.dat};
		\addplot[blue, dash dot, line width=1.5pt] file[] {tikz/data_1Dforward/1Dforward_p4_v1_t1.dat};
		\addplot[green, dotted, line width=1.5pt] file[] {tikz/data_1Dforward/1Dforward_p4_v2_t1.dat};
		\addplot[black, dashed] file[] {tikz/data_1Dforward/elBoundary1.dat};
		\addplot[black, dashed] file[] {tikz/data_1Dforward/elBoundary2.dat};
		\addplot[red] file[] {tikz/data_1Dforward/interface.dat};
		
		\legend{analytic, $\rho$-scaling, $c$-scaling, $\rho c$-scaling}
	\end{axis}
\end{tikzpicture}
			\vspace{-0.5cm}
			\caption{$t = \SI{0.5}{\second}$}
		\end{subfigure}
		%\hfill
		\begin{subfigure}[b]{1.0\textwidth}
			\centering
			\begin{tikzpicture}
	\begin{axis}[
		xmin = 0, xmax = 3,
		ymin = -0.5, ymax = 2.0,
		xtick distance = 0.5,
		ytick distance = 0.5,
		grid=both,
		width = \textwidth,
		height = 0.3\textwidth,
		xlabel = {$x$-coordinate},
		ylabel = {Displacement},
		legend style={at={(1,1)}, anchor=north east}]
		
		\addplot[red, line width=1.5pt] file[] {tikz/data_1Dforward/1Dforward_analytic_t3.dat};
		\addplot[black, dashed, line width=1.5pt] file[] {tikz/data_1Dforward/1Dforward_p4_v0_t3.dat};
		\addplot[blue, dash dot, line width=1.5pt] file[] {tikz/data_1Dforward/1Dforward_p4_v1_t3.dat};
		\addplot[green, dotted, line width=1.5pt] file[] {tikz/data_1Dforward/1Dforward_p4_v2_t3.dat};
		\addplot[black, dashed] file[] {tikz/data_1Dforward/elBoundary1.dat};
		\addplot[black, dashed] file[] {tikz/data_1Dforward/elBoundary2.dat};
		\addplot[red] file[] {tikz/data_1Dforward/interface.dat};
		
		\legend{analytic, $\rho$-scaling, $c$-scaling, $\rho c$-scaling}
	\end{axis}
\end{tikzpicture}
			\vspace{-0.5cm}
			\caption{$t = \SI{2.1}{\second}$}
		\end{subfigure}
		%\hfill
		\begin{subfigure}[b]{1.0\textwidth}
			\centering
			\begin{tikzpicture}
	\begin{axis}[
		xmin = 0, xmax = 3,
		ymin = -0.5, ymax = 2.0,
		xtick distance = 0.5,
		ytick distance = 0.5,
		grid=both,
		width = \textwidth,
		height = 0.3\textwidth,
		xlabel = {$x$-coordinate},
		ylabel = {Displacement},
		legend style={at={(1,1)}, anchor=north east}]
		
		\addplot[red, line width=1.5pt] file[] {tikz/data_1Dforward/1Dforward_analytic_t4.dat};
		\addplot[black, dashed, line width=1.5pt] file[] {tikz/data_1Dforward/1Dforward_p4_v0_t4.dat};
		\addplot[blue, dash dot, line width=1.5pt] file[] {tikz/data_1Dforward/1Dforward_p4_v1_t4.dat};
		\addplot[green, dotted, line width=1.5pt] file[] {tikz/data_1Dforward/1Dforward_p4_v2_t4.dat};
		\addplot[black, dashed] file[] {tikz/data_1Dforward/elBoundary1.dat};
		\addplot[black, dashed] file[] {tikz/data_1Dforward/elBoundary2.dat};
		\addplot[red] file[] {tikz/data_1Dforward/interface.dat};
		
		\legend{analytic, $\rho$-scaling, $c$-scaling, $\rho c$-scaling}
	\end{axis}
\end{tikzpicture}
			\vspace{-0.5cm}
			\caption{$t = \SI{3.5}{\second}$}
		\end{subfigure}
		%hfill
		\caption{1D interface problem $p = 4$}
		\label{1D_p4}
	\end{figure}
	
}

\section{2D forward problem}
\label{App:2D}
{
	\begin{figure}[H]
		\centering
		\begin{subfigure}[b]{0.3\textwidth}
			\centering
			\includegraphics[width=\textwidth]{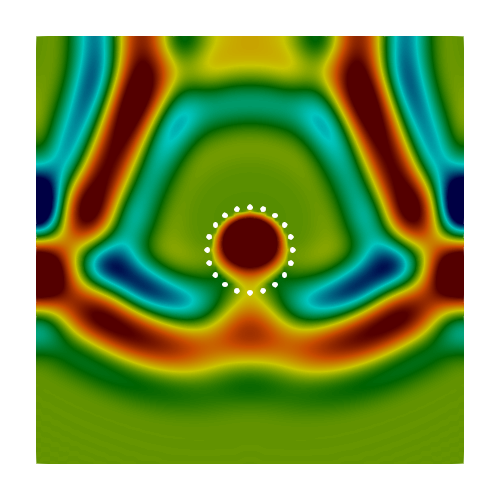}
			\caption{$\rho$-scaling}
		\end{subfigure}
		%\hfill
		\begin{subfigure}[b]{0.3\textwidth}
			\centering
			\includegraphics[width=\textwidth]{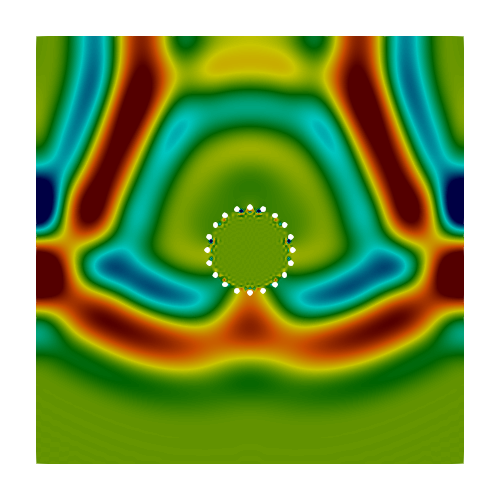}
			\caption{$c$-scaling}
		\end{subfigure}
		%\hfill
		\begin{subfigure}[b]{0.3\textwidth}
			\centering
			\includegraphics[width=\textwidth]{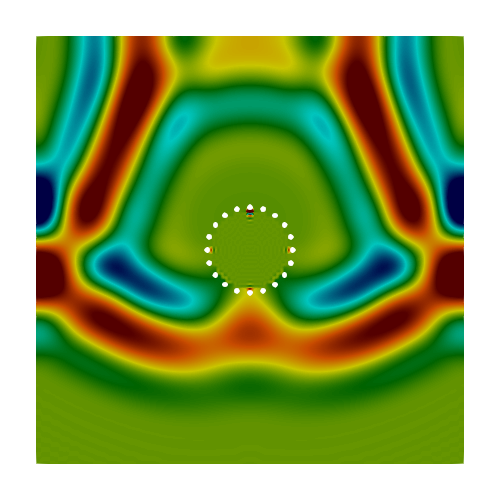}
			\caption{$\rho c$-scaling}
		\end{subfigure}
		\begin{subfigure}[t]{0.07\textwidth}
			\centering		
			\includegraphics{tikz/Colorbar_-0p01to0p01.tikz}
		\end{subfigure}
		%hfill
		\caption{2D plate with void $p=2$ at time $t = \SI{0.8e-05}{\second}$}
	\end{figure}

	\begin{figure}[H]
		\centering
		\begin{subfigure}[b]{0.3\textwidth}
			\centering
			\includegraphics[width=\textwidth]{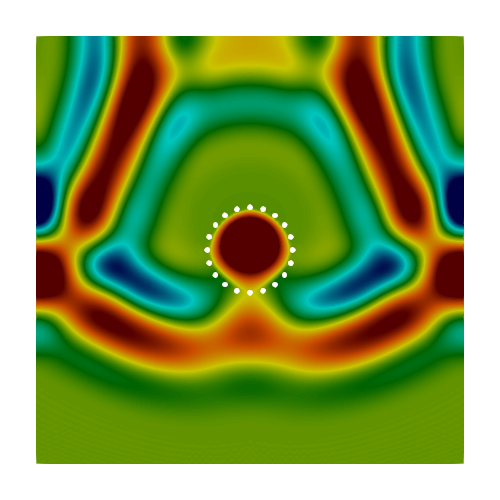}
			\caption{$\rho$-scaling}
		\end{subfigure}
		%\hfill
		\begin{subfigure}[b]{0.3\textwidth}
			\centering
			\includegraphics[width=\textwidth]{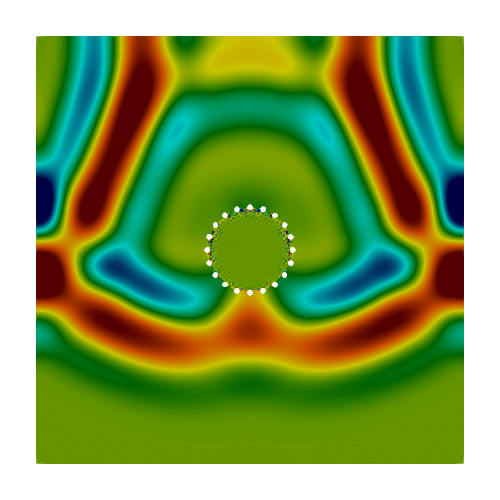}
			\caption{$c$-scaling}
		\end{subfigure}
		%\hfill
		\begin{subfigure}[b]{0.3\textwidth}
			\centering
			\includegraphics[width=\textwidth]{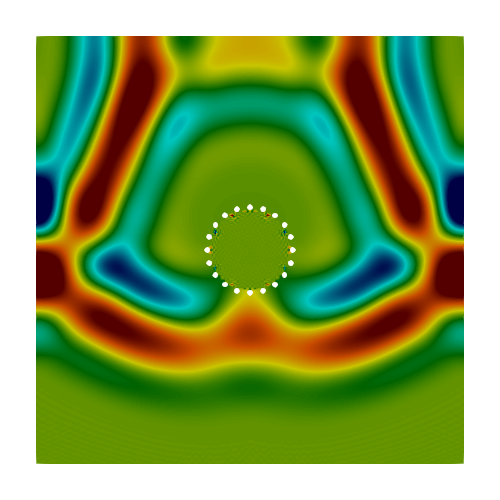}
			\caption{$\rho c$-scaling}
		\end{subfigure}
		\begin{subfigure}[t]{0.07\textwidth}
			\centering		
			\includegraphics{tikz/Colorbar_-0p01to0p01.tikz}
		\end{subfigure}
		%hfill
		\caption{2D plate with void $p=4$ at time $t = \SI{0.8e-05}{\second}$}
	\end{figure}
}
\section{2D inversion: idealized intermediate gradients}
\label{App:gradients}
{	

\begin{figure}[H]
	\centering
	\begin{subfigure}[t]{0.3\textwidth}
		\centering
		\includegraphics[width=\textwidth]{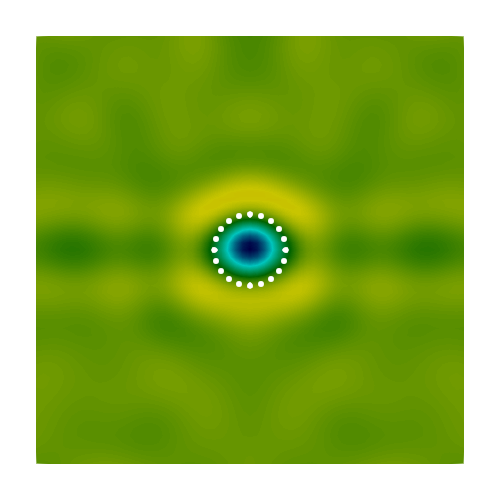}
		\caption{$\rho$-scaling}
	\end{subfigure}
	%\hfill
	\begin{subfigure}[t]{0.3\textwidth}
		\centering
		\includegraphics[width=\textwidth]{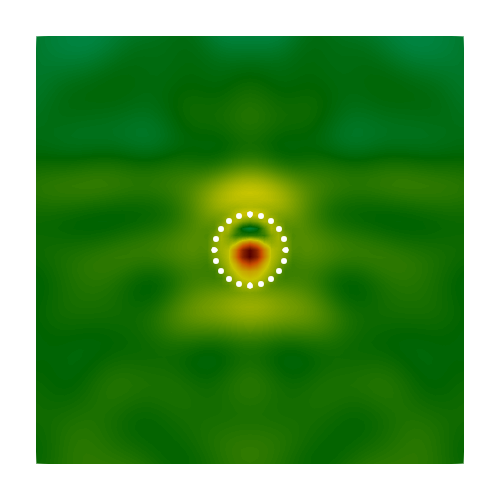}
		\caption{$c$-scaling}
	\end{subfigure}
	%\hfill
	\begin{subfigure}[t]{0.3\textwidth}
		\centering
		\includegraphics[width=\textwidth]{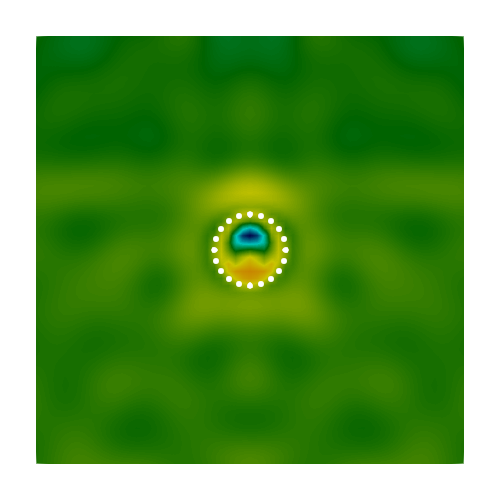}
		\caption{$\rho c$-scaling}
	\end{subfigure}
	%\hfill
	\begin{subfigure}[t]{0.07\textwidth}
		\centering
		\includegraphics{tikz/Colorbar_-1to1.tikz}
	\end{subfigure}
	%hfill
	\caption{Idealized intermediate gradients with $\gamma = 0.6$ inside the void}
	\label{gradient_0p6}
\end{figure}

\begin{figure}[H]
	\centering
	\begin{subfigure}[t]{0.3\textwidth}
		\centering
		\includegraphics[width=\textwidth]{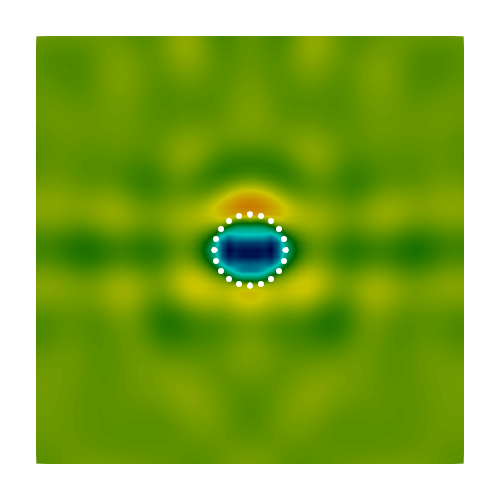}
		\caption{$\rho$-scaling}
	\end{subfigure}
	%\hfill
	\begin{subfigure}[t]{0.3\textwidth}
		\centering
		\includegraphics[width=\textwidth]{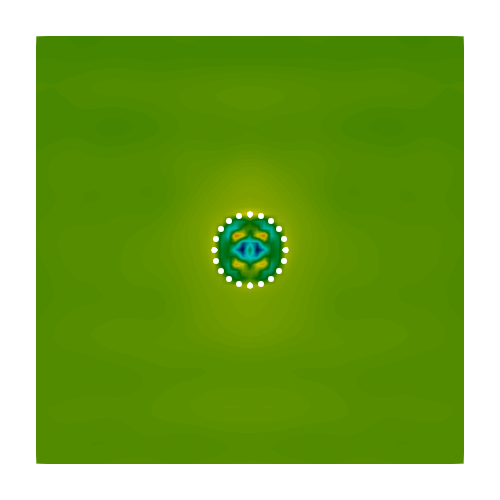}
		\caption{$c$-scaling}
	\end{subfigure}
	%\hfill
	\begin{subfigure}[t]{0.3\textwidth}
		\centering
		\includegraphics[width=\textwidth]{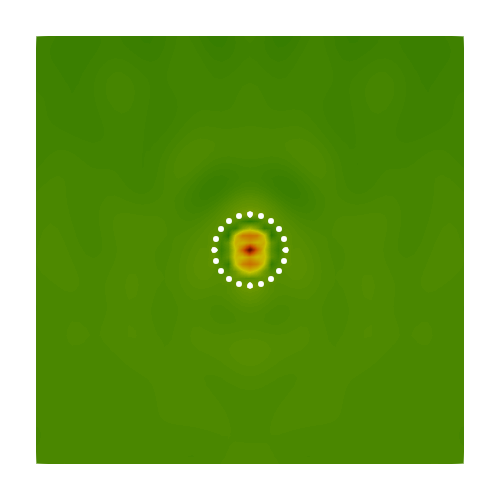}
		\caption{$\rho c$-scaling}
	\end{subfigure}
	%\hfill
	\begin{subfigure}[t]{0.07\textwidth}
		\centering
		\includegraphics{tikz/Colorbar_-1to1.tikz}
	\end{subfigure}
	%hfill
	\caption{Idealized intermediate gradients with $\gamma = 0.2$ inside the void}
	\label{gradient_0p2}
\end{figure}
}

% ------ Acknowledgements ------

\section*{Acknowledgements}
{
	We gratefully thank the Deutsche Forschungsgesellschaft (DFG) for their support support through the grants KO 4570/1-1 and RA 624/29-1.
}

% --------- References ---------

%\bibliographystyle{apalike}
\bibliographystyle{ieeetr}
%\bibliography{library}
%{\fontsize{10}{11.5}\selectfont\bibliography{library}}

\setlength{\bibsep}{3pt}
\setlength{\bibhang}{0.75cm}{\fontsize{9}{9}\selectfont\bibliography{library}}

\end{document}